\documentclass[sigconf]{acmart}

\AtBeginDocument{%
  }

\copyrightyear{2026}
\acmYear{2026}
\setcopyright{cc}
\setcctype{by}
\acmConference[ASSETS '26]{The 28th International ACM SIGACCESS Conference on Computers and Accessibility}{October 25--28, 2026}{Vila Nova de Gaia, Portugal}
\acmBooktitle{The 28th International ACM SIGACCESS Conference on Computers and Accessibility (ASSETS '26), October 25--28, 2026, Vila Nova de Gaia, Portugal}
\acmDOI{10.1145/3797867.3828996}
\acmISBN{979-8-4007-2521-0/2026/10}

\usepackage{tabularx} %
\usepackage{longtable} %
\usepackage{xcolor}
\usepackage[export]{adjustbox} %
\usepackage{makecell}
\usepackage{hyperref}
\usepackage{xspace}
\usepackage{multirow}

\usepackage[normalem]{ulem}
\usepackage{ifthen}

\usepackage{enumitem}

\usepackage{subcaption}

\newcommand{\tabref}[1]{Table~\ref{#1}}

\newcounter{ctr}

\newlength{\saveparindent}
\newlength{\saveparskip}
\definecolor{lgray}{gray}{0.85}

\newcommand{\imgborder}[2]{%
  \begingroup
    \setlength{\fboxsep}{0pt}%
    \setlength{\fboxrule}{1pt}%
    \color{lgray}%
    \fbox{\includegraphics[width=0.95\linewidth, valign=t]{#1}}%
  \endgroup
  \Description{#2}%
}

\begin{document}

\title[How Adults with Intellectual and Developmental Disabilities Navigate Security and Privacy Notifications]{``I’m trying not to get hacked:'' How Adults with Intellectual and Developmental Disabilities Navigate Security and Privacy Notifications}

\author{Hailey L. Johnson}
\email{hljohnson22@wisc.edu}
\orcid{0000-0003-1310-9948}
\affiliation{%
  \institution{Department of Computer Sciences\\University of Wisconsin--Madison}
  \streetaddress{Department of Computer Sciences, University of Wisconsin--Madison}
  \city{Madison}
  \state{Wisconsin}
  \country{United States}
}

\author{Julia Nonnenkamp}
\email{nonnenkamp@wisc.edu}
\orcid{0009-0006-4179-4224}
\affiliation{%
  \institution{Department of Computer Sciences\\University of Wisconsin--Madison}
  \streetaddress{Department of Computer Sciences, University of Wisconsin--Madison}
  \city{Madison}
  \state{Wisconsin}
  \country{United States}
}

\author{Bilge Mutlu}
\email{bilge@cs.wisc.edu}
\orcid{0000-0002-9456-1495}
\affiliation{%
  \institution{Department of Computer Sciences\\University of Wisconsin--Madison}
  \streetaddress{Department of Computer Sciences, University of Wisconsin--Madison}
  \city{Madison}
  \state{Wisconsin}
  \country{United States}
}

\author{Rahul Chatterjee}
\email{chatterjee@cs.wisc.edu}
\orcid{0000-0002-0308-8787}
\affiliation{%
  \institution{Department of Computer Sciences\\University of Wisconsin--Madison}
  \streetaddress{Department of Computer Sciences, University of Wisconsin--Madison}
  \city{Madison}
  \state{Wisconsin}
  \country{United States}
}

\renewcommand{\shortauthors}{Johnson et al.}

\begin{abstract}
Security and privacy notifications, such as login alerts, spam email warnings, and cookie consent requests, play a critical role in shaping users' responses to digital risks. Yet most notifications overlook cognitive accessibility, limiting their effectiveness for people with intellectual and developmental disabilities (IDD). We investigate how adults with IDD perceive and respond to common security and privacy notifications across mobile and web applications. Through a formative user study with seven adults with IDD, we identify three factors shaping understanding and decision-making: (1) interpretation is influenced by task and interface context; (2) unfamiliar terms, both technical and non-technical, are grounded in everyday concepts; and (3) uncertainty about outcomes leads to hesitation, avoidance, diagnostic exploration, or support-seeking. These findings lead to three design implications: (1) address context-dependent language misunderstandings beyond jargon simplification; (2) make action-outcome connections transparent; and (3) enable interdependent decision-making. Together, these insights aim to inform the design of more cognitively accessible security and privacy notifications that better support safe and supported user action.

\end{abstract}




\begin{teaserfigure}
  \includegraphics[width=\textwidth]{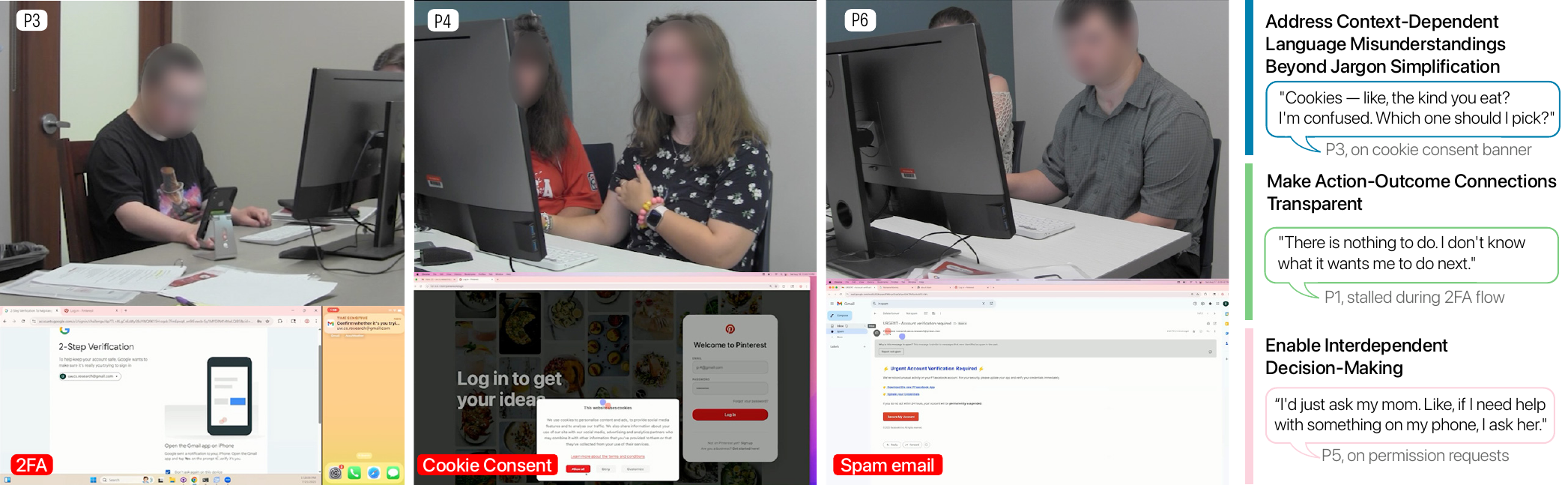}
  \Description{From left to right: P3 is interacting with a 2FA notification. P4 is interacting with a cookie consent notification. P6 is interacting with a spam email. All leading to three design implications. (1) Address context-dependent langauge misunderstandings beyond jargon simplification with supporting quote from P3 on cookie consent "Cookies? Like the kind you eat? I'm confused. Which one should I pick?" (2) make action-outcome connections transparent with supporting quote from P1 stalled during 2FA flow "There is nothing to do. I dont know what it wants me to do." (3) Enable interdependent decision making with supporting quote from P5 on permission requests "I'd just ask my mom. Like, if I need help with something on my phone, I ask her."}
  \caption{\textit{Overview of the Study Design and Design Implications.} This work examines how adults with intellectual and developmental disabilities interpret security and privacy notifications across web and mobile contexts. Through a qualitative study with seven participants (Left: P3, P4, and P6 interacting with three notifications), we identify factors that shape comprehension and decision-making. We derive three design recommendations (Right): (1) address context-dependent language misunderstandings beyond jargon simplification; (2) make action-outcome connections transparent; and (3) enable interdependent decision-making.}
  \label{fig:teaser}
\end{teaserfigure}

\maketitle

\newcommand{\spn}{S\&P notifications\xspace}
\section{Introduction}
\label{sec:intro}

Digital participation has become a societal norm, and this shift poses security and privacy (S\&P) risks to users. 
While most security protections operate invisibly in the background, critical points still demand user input, shifting the responsibility of decision-making onto individuals. 
These user-facing protections are not designed to accommodate the diverse ways people interpret and respond to them, particularly individuals with diverse cognitive profiles~\cite{wang2017third, renaud2022accessible, ma2013investigating}. With 1–3\% of the population living with an intellectual disability~\cite{patel2020clinical}, the design of S\&P protections must extend beyond technical safeguards to ensure inclusivity through design.

Responding to calls for inclusive S\&P~\cite{wang2017third, renaud2022accessible}, we situate our work in an emerging research direction that treats cognitive accessibility as a core ideal in security design by centering the perspectives of adults with intellectual and developmental disabilities (IDD). IDD is a condition defined by limitations in both \textit{intellectual functioning} (\textit{e.g.,} reasoning, problem solving, learning, communication) and \textit{adaptive behavior} (\textit{e.g.,} social, conceptual, and practical skills needed for daily life)~\cite{AAIDD, lee2019intellectual}.  Importantly, despite this clinical framing, individuals with IDD demonstrate diverse strengths and actively contribute to their families, communities, and workplaces~\cite{verdonschot2009community, mccausland2022potential}. As people engage with technologies across these contexts~\cite{johnson2024easy, morris2023examining, feng2008computer, jevne2022perspective}, ensuring that S\&P features are usable and understandable becomes essential to strengthening inclusion.

User-facing interventions often come in the form of \emph{S\&P notifications}, informational prompts that often require users to make decisions related to security or privacy. Users are expected to notice these notifications and take appropriate action, or no action if they believe there is no security or privacy concern 
~\cite{utz2023comparing,ndibwile2018comparative}. Despite their importance, many \spn pose persistent barriers to usability and inclusion~\cite{vance2019fog,li2016you}. Users regularly ignore important messages due to misunderstanding or alert fatigue~\cite{utz2023comparing,karyda2016data, reederExperienceSamplingStudy2018, bravo-lilloHarderIgnoreRevisiting2014}. To address these concerns, extensive research has investigated how to design more effective, attention-grabbing notifications to increase the adoption of protective measures.
~\cite{markertUnderstandingUsersInteraction2024,kitkowska2020enhancing,schaub2015design, gollaDriving2FAAdoption2021}. Still, many \spn require background knowledge, timed decisions, attention to technical details, or the ability to interpret lengthy legal or informational text -- demands that may not be appropriately designed for all users. Most prior work in usable S\&P, however, has focused on the general population, leaving the accessibility of these notifications for users with diverse cognitive needs underexplored~\cite{wang2017third, renaud2022accessible, ma2013investigating}. 

The minimal S\&P research conducted with individuals with IDD has largely focused on authentication methods, training interventions, and privacy tools, identifying both usability challenges and opportunities to support secure access. Usability challenges include the cognitive demands of managing complex passwords and the additional time required to complete logins~\cite{ma2013investigating, lazar2011understanding, furnell2021disadvantaged}. 
Other work has emphasized online safety training, privacy-preserving tools, and the important but complicated role of caregivers in mediating digital security~\cite{sundeepa2021effectiveness, venkatasubramanian2021exploring, chang2021towards, chalghoumi2019information, rocheleau2021caregivers}. Building on this foundation, this paper shifts attention from authentication and training to everyday S\&P notifications. We explore how adults with IDD understand, respond to, and evaluate these notifications \textit{in situ}. We aim to answer the following research question:

\begin{enumerate}\it
    \item[RQ:]  How do adults with IDD experience S\&P notifications, what challenges do they face, and what factors influence their understanding and decision-making?
\end{enumerate}
    
To answer the research question, we designed and conducted a formative user study with seven adults with IDD with two goals: (1) to understand their experiences with S\&P; and (2) to assess the usability of a range of S\&P notifications presented on browsers and smartphones. Qualitative analysis of the study data revealed three key themes that informed three design suggestions. Our key contributions include the following:
\begin{itemize}[leftmargin=*, itemsep=2px]
    \item \textit{Empirical insights into S\&P notifications with adults with IDD.} We report findings from a formative user study with seven adults with IDD, examining how they interpret and respond to common S\&P notifications across browser and smartphone contexts.  
    \item \textit{Strategies and barriers in notification use.} Through qualitative thematic analysis, we identify three themes: (1) notification interpretation was shaped by the ongoing task and interface context; (2) everyday concepts provided anchors for understanding and action; and (3) uncertainty about outcomes caused hesitation, avoidance, or support-seeking.
    \item \textit{Design suggestions for inclusive S\&P notifications.} Based on our findings, we propose three design suggestions: (1) address context-dependent language misunderstandings beyond jargon simplification; (2) make action-outcome connections transparent; and (3) enable interdependent decision-making. 
\end{itemize}

\section{Related Works}
\label{sec:relwork}
Our work builds on three intersecting strands of prior research: (1) the characteristics and diversity of adults with IDD, including variation in cognitive functioning, adaptive behaviors, literacy, and support needs; (2) usable security and privacy notifications, which have examined design effectiveness but often without attention to accessibility; and (3) online safety research for people with IDD. By drawing these strands together, we focus on how adults with IDD engage with everyday S\&P notifications.

\subsection{Characteristics and Diversity Within Intellectual and Developmental Disabilities}
\label{sec:relwork:IDD}
Intellectual and developmental disabilities (IDD) describe conditions characterized by limitations in intellectual functioning and adaptive behavior that originate before adulthood \cite{AAIDD, bartoshesky2021intellectual}. This community is heterogeneous, with independence and support needs influenced by co-occurring conditions such as communication disorders, mental health challenges, and age-related decline that further shape adaptive skills and technology use \cite{tasse2012construct, tasse2016relation}. Differences in attention and executive functioning also contribute to this variability, including documented challenges in working memory, processing speed, attentional control, and cognitive planning \cite{danielsson2010executive, hooper2018developmental, lanfranchi2010executive, bertelli2018intelligence, zagaria2021executive, deutsch2008attention}. Literacy and academic skills vary substantially across the IDD population, and limitations in reading comprehension and written expression can constrain access to education, employment, and independent living opportunities for some \cite{hua2019literacy, esbensen2010comparisons}.
This heterogeneity is reflected in the range of conditions that co-occur with IDD, including Down syndrome (DS), Autism Spectrum Disorder (ASD), Fragile X syndrome, Cerebral Palsy, Fetal Alcohol Spectrum Disorders, among others. People with IDD represent roughly 1–2\% of the U.S. population \cite{maulik2011prevalence, rzhetsky2014environmental}, with rates varying globally ($\sim$1-3\% \cite{patel2020clinical}) due to diagnostic and environmental differences. Within this broad group, DS and ASD partially illustrate the diversity of cognitive and adaptive profiles encompassed by IDD. DS is the most common single cause of intellectual disability, affecting about 1 in 1,499 individuals in the U.S. \cite{de2017estimation}. Approximately 1 in 3 children with ASD have co-occurring IDD \cite{shenouda2023prevalence}, and about 10–15\% of people with DS also have ASD, though overlapping developmental features often delay dual diagnosis \cite{dykens2007psychiatric, versaci2021down}. 

Despite this variability, adults with IDD demonstrate many strengths and actively participate in family, community, and employment contexts, especially when provided with appropriate supports \cite{verdonschot2009community, mccausland2022potential}.
Levels of independence vary, with some individuals living and working with minimal assistance and others relying on caregivers or community programs \cite{woolf2010adaptive}, making interdependence a significant feature of daily life \cite{reindal1999independence, bayor2021toward}. As digital technologies become central to work, education, and healthcare, adults with IDD may face both opportunities and barriers. They use mainstream and assistive tools to support daily living, but accessibility gaps persist, with rapidly evolving systems often outpacing adaptations \cite{glencross2021internet, johnson2023technology, johnson2024easy}.
With increasing use of technology also comes greater exposure to S\&P risks that demand rapid interpretation and decision-making. This study examines how adults with IDD engage with S\&P notifications to identify opportunities for inclusive security design.

\subsection{Usable Security and Privacy Notifications}
Much prior work has evaluated the design and effectiveness of security and privacy notifications: login and two-factor authentication alerts~\cite{gollaDriving2FAAdoption2021, markertUnderstandingUsersInteraction2024, redmilesShouldWorryCrossCultural2019, redmiles2017YouWant}, browser warnings against malware and encryption issues (\textit{i.e.,} using HTTP instead of HTTPS)~\cite{feltImprovingSSLWarnings2015, feltExperimentingScaleGoogle2014, akhaweAliceWarninglandLargescale2013, bravo-lilloYourAttentionPlease2013, egelmanYouveBeenWarned2008, almuhimediYourReputationPrecedes2014, reederExperienceSamplingStudy2018}, spam or phishing email warnings~\cite{petelkaPutYourWarning2019, sharevski2024AssessingSuspicious}, and data-use consent pop-ups~\cite{utz2019UninformedConsent, nouwens2020DarkPatterns}. 
Researchers have sought to improve notification interfaces through participatory design~\cite{redmiles2017YouWant, gorski2020ListenDevelopers}, modeling how users understand severity~\cite{redmilesShouldWorryCrossCultural2019, mendel2022ProactiveSupporta}, and leaning on semantics and semiotics (\textit{e.g.,} the color red) ~\cite{sunshineCryingWolfEmpirical2009, bravo-lilloYourAttentionPlease2013, wogalter2002ResearchbasedGuidelines}. 
Still, most quantitative studies of user behavior have found that users dismiss \spn more often than would be desired. Some~\cite{sunshineCryingWolfEmpirical2009, egelmanYouveBeenWarned2008} attributed this to \textit{habituation}, when end-users become less responsive to notifications after encountering them frequently~\cite{kim_wogalter_habituation_2009} or misunderstandings about the warning's meaning or the underlying system mechanism~\cite{reederExperienceSamplingStudy2018, bravo-lilloBridgingGapComputer2011}. When deploying privacy notifications such as cookie consent banners, websites often use deceptive language and visual cues ("dark patterns") to nudge users toward choosing permissive settings~\cite{nouwens2020DarkPatterns, utz2019UninformedConsent, brignullDP2023, gray2018DarkPatterns}. 

Despite extensive notification research, \textbf{accessibility of \spn for users with IDD remains understudied.} 
However, incorporating cognitive adaptations in non-security notifications is demonstrably beneficial~\cite{rehm2024notification, jamieson2020technology, eskandar2022designing}. For example, \citet{eskandar2022designing} designed a reminder system for individuals with mild cognitive impairment in automated vehicles, showing that multimodal cues, predictable timing, and adaptive prompting improved comprehension and reduced anxiety. Still, cognitive accessibility for \spn is underexplored. Existing work in this niche focuses on S\&P social support and information-seeking primarily among older adults~\cite{mendel2022ProactiveSupporta, nicholson2019importantHeadline}; \citet{mendel2022ProactiveSupporta} explored how interpersonal help during ``supportable moments'' can help older adults resolve stressful security messages.
Our study seeks to understand if \spn surface unique challenges for adults with IDD.

\begin{table*}[t]
    \caption{Taxonomy of Security Interventions and Intrusions~\cite{addasPinsonneault2015}, with examples: \textnormal{\textit{Left}: Actionable Intervention --- 2FA prompt during web login to Gmail. The notification relates to the primary task (logging in) and requires input. \textit{Middle}: Informational Intervention --- While the user browses the web (primary task), Chrome informs the user if malware is found in a download. \textit{Right}: Intrusion --- Pinterest login alert encountered during an unrelated primary task (reading email). This notification would be \textit{actionable} if it prompted the user to approve a login request; otherwise, \textit{informational}.}}
    \Description{A three-column table showing examples of different types of security and privacy notifications. The first column, Actionable Intervention, shows a Gmail two-factor authentication prompt. The second column, Informational Intervention, shows a Chrome browser warning about a dangerous file download. The third column, Intrusion, shows a Pinterest login email only visible in Gmail.}
    \label{tab:taxonomy}
  \centering
  \setlength{\tabcolsep}{10pt}    
  \begin{tabularx}{\textwidth}{
      >{\centering\arraybackslash}X
      >{\centering\arraybackslash}X 
      >{\centering\arraybackslash}X
    }
    \toprule
    \textbf{Actionable Intervention} & \textbf{Informational Intervention} & \textbf{Intrusion} \\
    \midrule
    
    {\small 2-factor authentication prompt during login (primary task).} & 
    {\small Dangerous download warning while browsing (primary task).} & 
    {\small New login email, which could be actionable or informational.} \\

    \imgborder{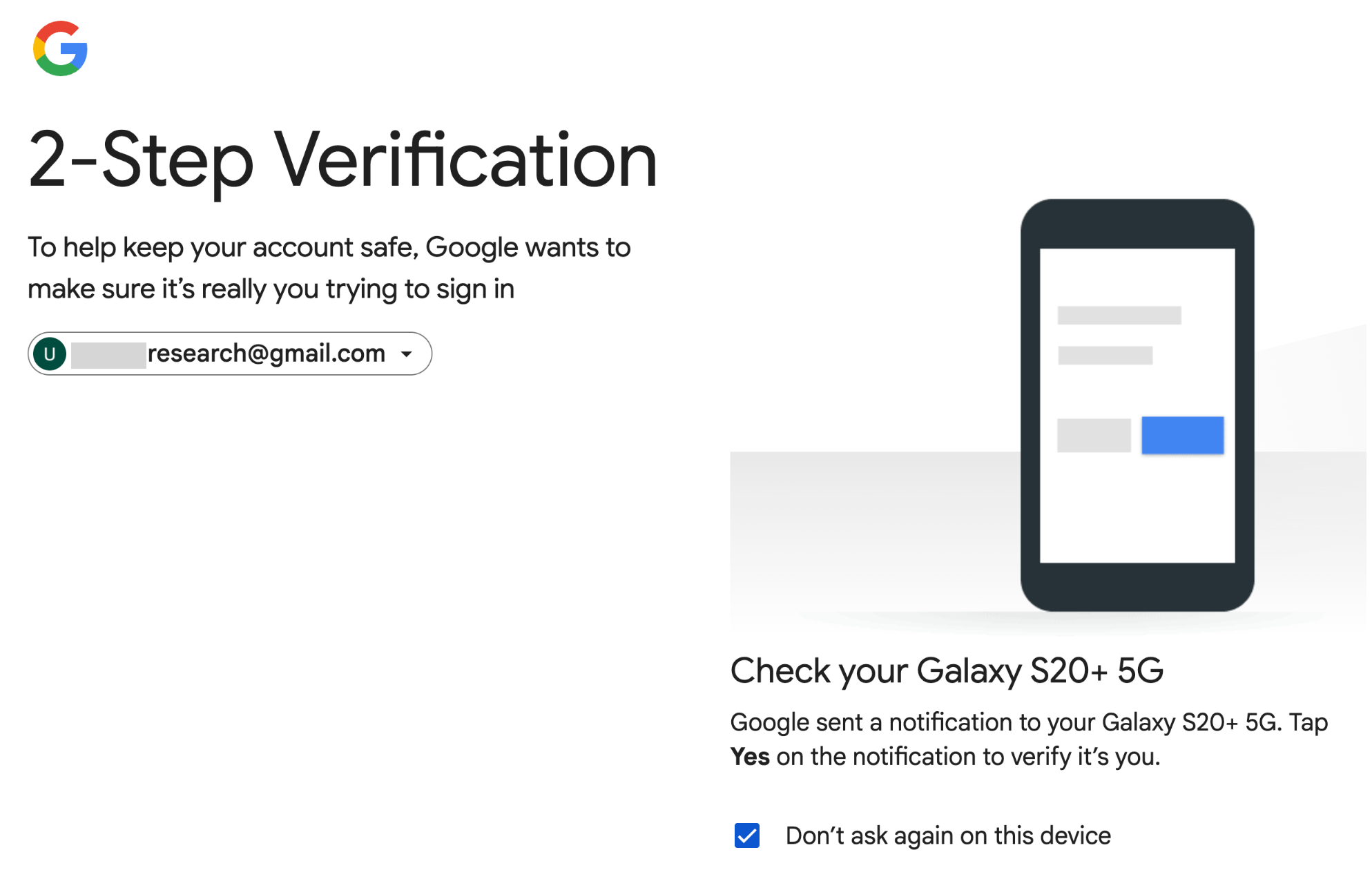}{Screenshot of a Google UI with two columns. On the left, note text saying ``2-Step Verification. To help keep your account safe, Google wants to make sure it's really you trying to sign in.''. On the right, there is an illustration of the Gmail app opening on a phone. Below the image there is text: "Check your Galaxy S20+5G. Google sent a notification to your Galaxy S20+5G. Tap YES on the notification to verify it's you" and a checkbox next to text "Don't ask again on this device.} 
    &
    \imgborder{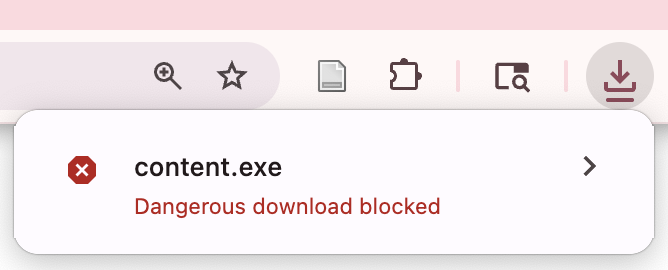}{Screenshot of a pop-up warning that appears in the top right of the Chrome browser near the downloads tab. It is a white rectangle with an icon on the left (red octagon icon with a white X) with the file name "content.exe" "Dangerous download blocked" in the middle. On the right, there is an arrow that expands the pop-up with more informational links.}
    &
    \imgborder{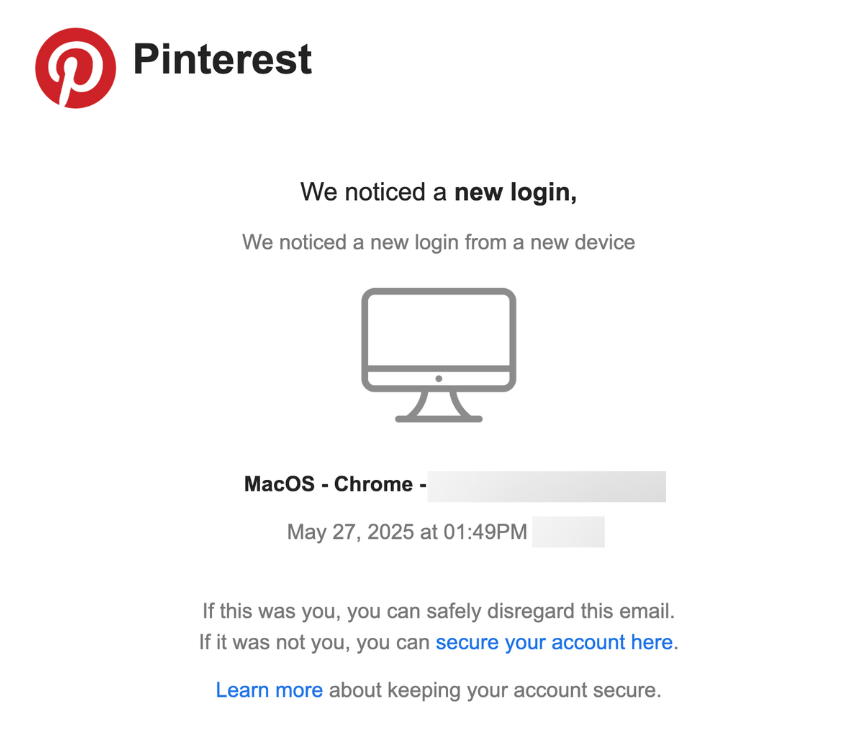}{This is an email notification. Within the email you see in the top left "Pinterest" and their logo. Reading from top to bottom it says "We noticed a new login." Below that, it has an image of a computer screen. Below that, it has device information in bolded font (MacOS, Chrome, [redacted location]. Below that, it has the date and time of the notification with [redacted time zone]. Below that, it explains that if it was not you, you should secure your account, which is a hyperlink. And finally, at the bottom is a learn more hyperlink about how to keep your account secure.} 
    \\

    \addlinespace[10pt] 
    
  \end{tabularx}
      
\end{table*}

\subsection{Security and Online Safety for People with Intellectual and Developmental Disabilities}
Individuals with IDD remain largely underrepresented in S\&P research, though recent work calls for making these systems more accessible. \citet{wang2017third} frames this as a ``third wave'' of S\&P research that moves beyond traditional user-centered security research toward approaches that center diverse human abilities. Likewise, \citet{renaud2022accessible} argues that accessibility should stand alongside security, privacy, and usability as a core design dimension, as inaccessible systems exclude both people with disabilities and users who struggle with S\&P interactions.
Prior work has shown that individuals with IDD and their support people manage everyday security and privacy collaboratively. \citet{venkatasubramanian2021exploring} found participants frequently shared PINs or passwords with relatives. Similarly, \citet{chalghoumi2019information} observed that participants were eager to use new technologies but often disclosed personal information or credentials without recognizing associated risks, noting that caregivers' instruction influenced security behavior. From the caregiver perspective, strategies for balancing autonomy and safety included limiting data access, guiding disclosure of sensitive, and providing just-in-time feedback ~\cite{rocheleau2021caregivers,si2025digital}. 

Beyond interdependent practices, prior work has examined how adults with IDD engage with authentication tools, such as PINs, passwords, and biometrics \cite{venkatasubramanian2021exploring, chalghoumi2019information, ma2013investigating, lazar2011understanding, furnell2021disadvantaged}. \citet{ma2013investigating} showed that adults with DS could successfully create and use alphanumeric passwords, though this took longer compared to neurotypical users (indicating a difference in processing speed) and improved with familiarity (indicating learnability). Similarly, \citet{lazar2011understanding} found that experienced users with DS could manage multiple passwords and complete CAPTCHA tests. Despite these successes, password systems remain cognitively demanding \cite{furnell2021disadvantaged, furnell2022accessible}, and empirical studies with adults with DS report longer task times, higher cognitive load, and difficulty with multi-step interactions~\cite{ma2013investigating, lazar2011understanding}. As described in Section~\ref{sec:relwork:IDD}, these patterns are consistent with differences in processing speed, working memory, and attentional control in IDD populations. 

Researchers often recommend multimodal and accessible options to reduce memory and attentional load \cite{furnell2021disadvantaged, furnell2022accessible}. \citet{kavrestad2021usable} similarly emphasize design principles such as text-to-speech functionality and conciseness. 
Beyond system design, researchers have examined educational and caregiver interventions. Gamified cybersafety programs that addressed scams and cyberbullying also improve basic awareness  \cite{sundeepa2021effectiveness}. \citet{si2025digital} found that parents and teachers of young people with IDD can use scenario-based simulations, multi-sensory cues, and gamified activities to teach digital safety. While prior work shows that security systems (primarily authentication) are learnable with time and support, our study turns to everyday S\&P notifications, examining how adults with IDD perceive, interpret, and act on them in the moment.

\section{Method}
\label{sec:methods}
This study examined how adults with IDD experience and interact with everyday S\&P notifications. We conducted a formative user study with seven adults with IDD, who encountered realistic notifications integrated with semi-structured interviews. Because S\&P notifications often appear as secondary tasks, each study task was designed around a primary activity (\textit{e.g.,} checking email, browsing a social media website, using a mobile app) with embedded notifications that requested a decision or response. This let us capture natural reactions to notifications without making security the explicit focus of the task. In this section, we explain (1) how we selected security and privacy notifications to evaluate; (2) participant demographics; (3) outline the study protocol; (4) our inclusive and ethical study design choices; and (5) the data collection and analysis procedures.

\subsection{Selecting Notifications}
\label{sec:methods:defn}

\citet{mccrickard2003ModelNotification} define \textit{notification systems} as interfaces ``attempting to deliver current, valued information through a variety of platforms and modes'' such as operating systems, browsers, and mobile applications.
\textit{Security and privacy (S\&P) notifications} deliver security- or privacy-related information, and they often additionally require implicit or explicit decision-making~\cite{mendel2022ProactiveSupporta}.
Privacy notifications include cookie consent banners, advertising or tracking notices, and data-sharing requests. Security notifications include browser warnings (\textit{e.g.,} when using HTTP instead of HTTPS), unfamiliar sign-in alerts, and software update reminders.

To ensure users encountered a diverse set of \spn, we selected multiple from the ``interruption'' categories identified by \citet{addasPinsonneault2015}.\footnote{We disregard ``system'' notifications ~\cite{addasPinsonneault2015} as it was only possible to include notifications triggered by human actions.}
Their taxonomy has two dimensions: \textit{Content relevance} describes whether the notification is relevant to the user's primary task (\textit{intervention}) or interrupts the primary task with other information  (\textit{intrusion}). \textit{Content structure} describes whether a notification is \textit{informational} or \textit{actionable}. In this context, we take \textit{actionable} to mean it requires the user to take security- or privacy-related action to proceed with their task (\textit{e.g.,} approving a login request).
Examples are shown in \tabref{tab:taxonomy} and Appendix~\ref{app_notifs}.

\subsection{Participant Recruitment and Demographics}
\label{sec:methods:demographics}

We recruited participants through email lists of prior participants who had opted in to future contact and flyers distributed to community organizations in our region and neighboring states. These organizations included assisted-living programs, Down syndrome groups, ASD organizations, and broader cognitive disability service providers. Three individuals withdrew due to unrelated scheduling constraints. Recruitment occurred over four months, resulting in a final sample of seven adults.

Table~\ref{tab:Demo} shows the demographic information for our seven participants, aged 19--42 years ($M=27.7$, $SD=9.18$; four women, three men). Five participants had a primary diagnosis of Down syndrome, while one participant had a primary diagnosis of autism spectrum disorder (ASD), and one had a primary diagnosis of IDD. Additional reported comorbidities included hearing and vision loss, dysarthria, dyspraxia/apraxia of speech, sensory processing disorder, seizure disorder, and leukemia remission. The IDD population is broad and diverse. Our participant pool does not capture this full heterogeneity; however, it reflects well-studied diagnostic groups that illustrate distinct cognitive and adaptive profiles relevant to understanding how adults with IDD engage with S\&P notifications \cite{maulik2011prevalence, shenouda2023prevalence, dykens2007psychiatric}.
Each participant was informed that they could invite a support person (\textit{e.g.,} parent, sibling, caregiver) to increase comfort during the study, and four participants opted to do so. The demographic information for these support persons is provided in Table~\ref{tab:Demo}, with ages ranging from 56--70 years ($M=60.9$, $SD=6.29$; four women).

\begin{table*}[t]
\centering\small
   \Description{Demographic characteristics of seven adult participants with intellectual and developmental disabilities (IDD) and their support persons. Participants (P1–P7) range in age from 19 to 42. Most participants are White (5 of 7), with one Asian participant and one Black or African American participant. Gender includes four women and three men. Education levels include being in or having completed transition programs, some college, or high school graduation. Common reported comorbidities include Down syndrome (5 participants), ASD (1), hearing loss (4), vision loss (2), sensory processing disorder (3), seizure disorder (1), leukemia remission (1), dysarthria (1), and dyspraxia/apraxia of speech (1). Four participants (P1, P3, P4, P6) had support persons present; three (P2, P5, P7) did not. All support persons were women with at least a bachelor’s degree. Support persons ranged in age from 50 to 65+ and were primarily White, with one identified as Asian/White.}
   \caption{Demographic characteristics of adult participants with IDD and their support persons, if present. \textnormal{All support persons were women with a Bachelor's degree or above. Down syndrome (DS), Autism Spectrum Disorder (ASD).}}
    \label{tab:Demo}
    \begin{tabular}{ccp{1.2cm}lp{3.3cm}p{5cm}p{3.2cm}}
    \toprule
        \textbf{PID} & \textbf{Age} & \textbf{Race} & \textbf{Gender} & \textbf{Education} & \textbf{Reported Comorbidities} & \textbf{Support Person}\\
        \midrule
        P1 & 21-30 & Asian & Man & Some college & DS, Hearing loss & SP1 (50-64; Asian, White; Graduate degree)\\
        P2 & 31-40 & White & Woman & Finished transition program & DS, Hearing loss& -\\
        P3 & 31-40 & White & Man & Finished transition program & DS, Vision \& Hearing loss, Sensory processing disorder & SP3 (65+; White, Bachelor degree)\\
        P4 & 18-20 & White & Woman & In transition program & ASD, Leukemia Remission, Seizure Disorder& SP4 (50-64, White, Bachelor degree)\\
        P5 & 18-20 & White & Woman & In transition program & DS, Vision \& Hearing loss & -\\
        P6 & 18-20 & White & Man & In transition program & DS, Dysarthria, Dyspraxia\slash Apraxia of speech, Sensory Processing Disorder & SP6 (50-64; White, Graduate degree)\\
        P7 & 41-50 & Black or African American & Woman & Graduated high school & Sensory processing disorder& -\\
        \bottomrule
    \end{tabular}
 
\end{table*}

\subsection{Study Protocol}
\label{sec:methods:protocol}

This qualitative study integrated semi-structured interviews and behavioral analysis from usability testing to explore how adults with IDD experience and interpret S\&P notifications. Regardless of the presence of support people, participants with IDD were encouraged to be the primary decision-makers and to complete tasks as independently as possible. The Institutional Review Board at the authors’ university approved the study.

The study consisted of four main phases: (1) \textit{online or in-person screening and consent}; (2) \textit{semi-structured interview to assess general technology and computer security experience}; (3) \textit{usability testing through three tasks using a lab-provided computer and smartphone}; and (4) \textit{post-study interview and payment}. The initial screening included eligibility questions, rapport-building icebreakers, and a walkthrough of an accessible consent form segmented into short sections with supportive visuals. Once consent was obtained, participants scheduled an in-person study session. Screening and consent took 15-30 minutes, and the in-person study lasted 1-1.25 hours. If participants invited support persons and they were present, they were asked to complete a separate consent process. Study sessions were held at the authors' university or in alternative accessible locations (\textit{e.g.,} community centers, libraries) at participants' request. Breaks were offered at any time. At the conclusion of all tasks, a short post-study interview was conducted to reflect on the tasks, interface designs, and broader preferences for online safety tools. Participants were compensated \$20 an hour for their time. Study materials are available on OSF.\footnote{The protocol is available at the OSF Link: \url{https://osf.io/6xqvn/overview?view_only=575d37e431b749df9174e075614d4fcf}}

\begin{table*}[t]
\caption{Overview of the three study tasks, their subtasks, and S\&P notifications participants interacted with during the study tasks. \textnormal{Three example notifications are highlighted. The notification set is in Appendix~\ref{app_notifs}.}}
\label{tab:task_notifs}
\Description{A 2-column, 3-row table showing the 3 tasks participants completed alongside a screenshot of an example notification from each task.}

\centering
\setlength{\tabcolsep}{18pt}
\setlength{\aboverulesep}{6pt}
\setlength{\belowrulesep}{6pt}
\begin{tabular}{p{0.52\textwidth} p{0.33\textwidth}}
\toprule
\textbf{Tasks \& Steps} & \textbf{Example Notification} \\
\midrule

\textbf{Task 1: Email} & {} \\
\begin{enumerate}[itemsep=2px, leftmargin=*]
    \item Log in on a computer, encountering instruction screen \textbf{[2FA-C]}.
    \item Authenticate via phone, encountering push banner \textbf{[2FA-B]} and in-app approval \textbf{[2FA-I]}.   
    \item Browse marketing emails.
    \item Locate an email in Spam folder with the gray Spam banner \textbf{[SP-G]}.
\end{enumerate}
& 
\setlength{\fboxsep}{0pt}%
\setlength{\fboxrule}{1pt}%
\color{lgray}%
\fbox{\includegraphics[width=\linewidth, height=4cm, keepaspectratio, valign=t]{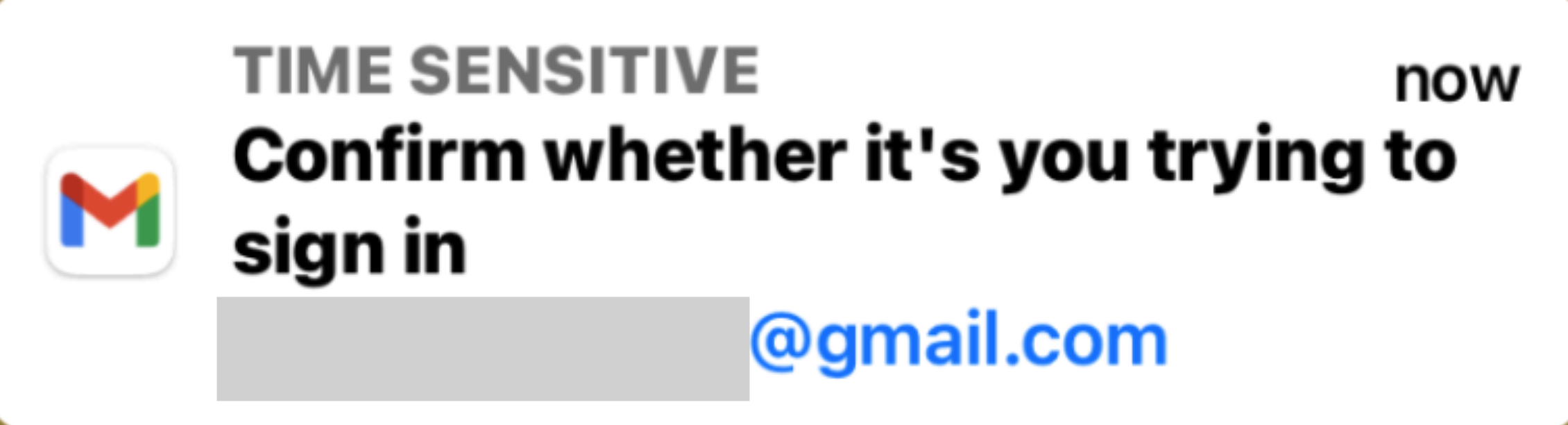}}%
\color{black}%
\Description{Screenshot of a push notification banner from Gmail saying "Time sensitive. Confirm whether it's you trying to sign in [redacted email address]" with Gmail's logo (color envelope) on the left.}
\newline\newline \small \textbf{2FA-B:} Push banner \\
\midrule

\textbf{Task 2: Social Media} & {} \\
\begin{enumerate}[itemsep=2px, leftmargin=*]
    \item Log in on a computer and accept the Cookie Consent prompt \textbf{[CC]}.
    \item Browse posts and click external links, encountering warnings for malicious website \textbf{[MW]}, dangerous download \textbf{[DD]}, and invalid HTTPS \textbf{[HT]}.
    \item Return to Gmail (Task 1) to inspect new sign-in email alerts \textbf{[NSI]}.
\end{enumerate}
& 
\includegraphics[width=\linewidth, height=4.2cm, keepaspectratio, valign=t]{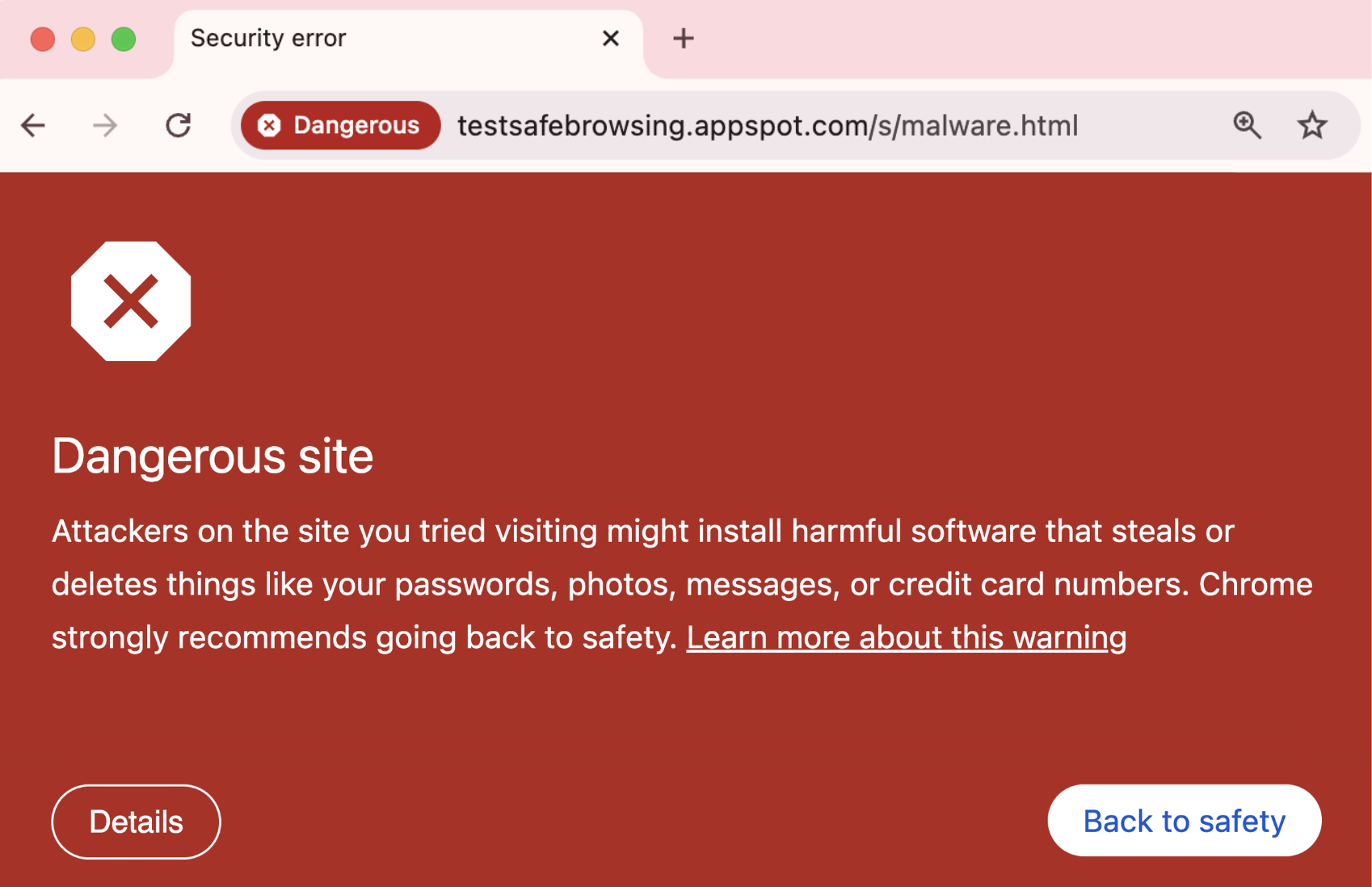}
\Description{Screenshot of bright red Chrome browser warning taking up the whole screen. At the top, there is a white octagon icon with a red X in the middle with large white font "Dangerous site" below it. The paragraph below says:  "Attackers might be trying to steal your information from expired.badssl.com (for example, passwords, messages, or credit cards)." with a hyperlink stating "Learn more about this warning." Below this there are two oval buttons: "Advanced" which is the same color as the background and unemphasized, and "Back to Safety," which is blue and visually prominent.}
\newline \newline \small \textbf{MW:} Chrome malicious site warning \\
\midrule

\textbf{Task 3: Weather App} & {} \\
\begin{enumerate}[itemsep=2px, leftmargin=*]
    \item Launch app, encountering screen recording \textbf{[SRI]} and microphone use \textbf{[MUI]} indicator dots.
    \item Click through permission pop-ups: cross-app tracking \textbf{[TP]}, location requests \textbf{[LP]}, and standard notification requests \textbf{[NP]}.
    \item Locate and report the current weather.
\end{enumerate}
& 
\includegraphics[width=\linewidth, height=4.2cm, keepaspectratio, valign=t]{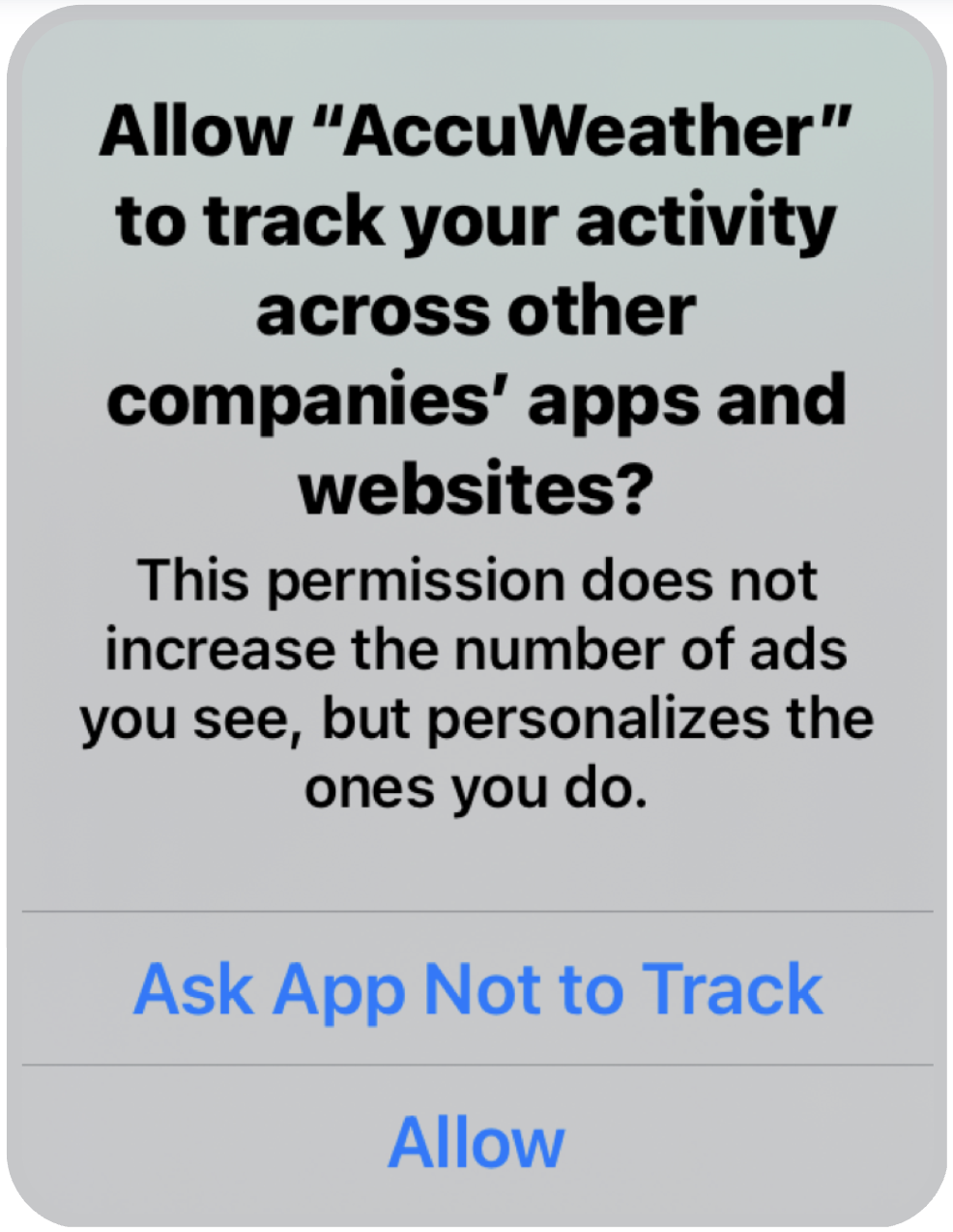}
\Description{Screenshot of an iOS notification pop-up requesting permission to ``Allow "AccuWeather" to track your activity across other companies and websites?'' Underneath it says ``This permission does not increase the number of ads you see, but personalizes the ones you do.'' Underneath it has two buttons, The top one says ``Ask app not to track'' and the bottom one says ``Allow.'' The notification appears in the middle of the iPhone screen.}
\newline \newline \small \textbf{TP:} Cross-app Tracking Permission \\
\bottomrule
\end{tabular}
\end{table*}

\subsubsection{Task Design}
\label{sec:methods:taskdesign}

Participants completed three tasks using a smartphone and a computer provided for the duration of the study. Depending on the participant's preferred operating system, they used an iPhone 7 (iOS 15) or a Samsung Galaxy S20 (Android 13), and a 2020 MacBook Air (macOS Monterey) or a Lenovo ThinkPad T490 (Windows 11). 
Participants accessed the computer with an external monitor, keyboard, and mouse, and interviewers adjusted the text size to participants' preferences. The smartphone was placed on a vertical stand beside the monitor, and participants could pick up the phone as desired. Participants used Chrome for all web browser tasks due to its dominant market share~\cite{wikimedia_browser_dashboard}. After each task, the interviewers reviewed all notifications seen by participants using paper printouts of each variant (see Appendix~\ref{app_notifs}) and prompted feedback on comprehension, design elements, and ease of use.

\paragraph{Task 1: Email Navigation and 2FA (Computer + Smartphone)} 
Participants signed in to a Gmail account with a saved username \& password on the computer, and then completed a two-factor authentication (2FA-C) using Gmail's mobile app (2FA-I, 2FA-B). They explored an inbox containing 10 personalized marketing emails and a simulated phishing email in the spam folder, while prompted to reflect on spam banners (SP-G) and suspicious content. 

\paragraph{Task 2: Pinterest-like Website (Computer)} 
Participants interacted with a custom Pinterest-style image-board interface tailored to their interests. A cookie consent prompt (CC) appeared on entry, and users logged in using a saved username and password. Opening image posts redirected participants to external websites where they encountered three browser warnings (MW, DD, HT). Then, participants returned to the email inbox used in Task 1 to review two new sign-in (NSI) email alerts triggered by the initial site ``login'' --- one with correct user details (time, location, operating system, browser) and another with incorrect, or fraudulent, details. The interviewers then prompted participants to differentiate between the two emails and explain how they identified potential risks.

\paragraph{Task 3: Weather App Permissions (Smartphone)} In this task, participants used AccuWeather~\cite{AccuWeatherApp} to identify the current temperature. Upon opening the app, which had been freshly reinstalled before each study, typical mobile notifications were triggered, such as terms of service agreements and permission request pop-ups (TP, LP, and NP). The researchers had additionally initiated screen and audio recording in the background, which caused indicators (SRI, MUI; typically small dots in the top corner of the screen) to appear.

\subsection{Inclusive Methods and Ethical Study Design}
\subsubsection{Inclusive Methods}
\label{sec:methods:inclusive}
We incorporated a range of practices drawn from prior literature on inclusive study environments \cite{l2025accessible}. Participants could include a support person for comfort, though we directed questions only to the participant with IDD to preserve their role as primary decision-maker. Study materials were personalized to participants' interests based on the icebreakers: inbox messages came from familiar organizations, Pinterest-style images matched their hobbies, and participants chose their preferred device (Mac vs.\ Windows, iPhone vs.\ Android). When participants showed avoidant behaviors, researchers offered guiding questions (``What do you think?''), reassurance (nods, smiles), or hints (``You could open the email and look if you want.''), always letting participants act first and using prompts to encourage exploration rather than signal a ``correct'' choice. Visual aids were used to support comprehension and reflection, including printed copies of each notification and an adapted consent form split into short, illustrated sections. Participants could also choose their study location (\textit{e.g.,} our university or a local library), and breaks were offered between tasks.

\newcommand{\ethicalstudydesignchoices}{Ethical Study Design Choices}
\subsubsection{\ethicalstudydesignchoices}
\label{sec:methods:ethical}

This study exposed users to simulated cybersecurity risks (\textit{e.g.,} phishing emails, malicious websites) and prompted them to make choices when faced with such risks. Moreover, individuals with IDD are often considered \textit{at-risk users} because unmet accessibility needs can increase their vulnerability to cyberattacks or result in disproportionate harm~\cite{matthews2025SupportingDigitalSafety, warford2022AtRiskUserResearch, bellini2024SoKSafer}. 
Although participants did not use their personal devices, accounts, or information for the tasks, we took additional steps to prevent exposure to real malicious content, emotional distress, and negative training effects~\cite{distler2021SystematicRisk, bellini2024SoKSafer}.
While we used genuine marketing emails from companies and organizations in Task 1, we blocked navigation from these emails to external links. We designed the phishing email to trigger Gmail's spam filter through known signatures~\cite{dewan2014Stylometric} without malicious content. Similarly, Chrome's website warnings in Task 2, while authentic, were triggered by benign security-testing websites.\footnote{BadSSL (\url{https://badssl.com})~and Google's official test site  (\url{https://testsafebrowsing.appspot.com}).}

Only one participant made an explicitly unsafe choice by clicking a link in the simulated phishing email. To avoid negative training effects, the interviewers reviewed the participant's choice with their support person. They had a post-task follow-up conversation about how to mitigate this risk going forward.

\subsection{Data Collection and Analysis}

\subsubsection{Data Collection}
\label{sec:methods:collection}

Data were collected through a combination of audio and video recording, researcher observations, eye-tracking, and system logging, supported by \citet{hollomotz2018successful}. Semi-structured interviews were conducted before, between, and after tasks to capture participants’ perspectives on technology use, experiences with S\&P in daily life, their understanding of notification purposes, and the perceived importance of different cues. These interviews also explored the role of interdependence~\cite{bennett2018InterdependenceFrame} in online safety. Field notes, taken primarily by the second author, supplemented interviews and recordings by documenting participant behaviors, navigation choices, and task sequences. Researchers additionally noted the amount of assistance required and visible signs of confidence, frustration, or confusion to contextualize decision-making. Eye-tracking data were collected using a Tobii Pro Fusion eye tracker to examine attention patterns. Calibration proved challenging for some participants with Down syndrome due to eyesight and difficulty maintaining focus, resulting in incomplete data for three participants.
Finally, computer interactions were logged at the device level by recording click data to trace user choices.

\subsubsection{Data Analysis}
\label{sec:methods:analysis}

We employed Reflexive Thematic Analysis (RTA) \cite{braun2012thematic, braun2021can} to explore how adults with IDD interact with S\&P notifications. RTA supports inductive pattern-finding that moves from semantic (explicit, observable) to latent (implicit, underlying) meaning, treating the researcher's interpretation as central to theme development \cite{braun2021can}.
Our process followed five stages: (1) conducting the study and collecting data; (2) transcribing recordings and assembling composite videos that synchronized room footage, participant audio, and screen recordings with click and eye-tracking visualizations when available; (3) one researcher completing an initial round of semantic coding of participant behaviors (\textit{e.g.,} interaction sequences, verbal statements, eye-tracking, \textit{etc.}) followed by a second researcher coding random subsets of each participant's transcript and composite video to develop shared analytic familiarity with the dataset; (4) jointly developing initial themes through reflexive engagement with qualitative codes, field notes, and researcher memos recorded throughout data collection and analysis; and (5) repeating coding and theme development in two subsequent iterations.
Researchers coded transcripts alongside composite recordings, allowing behavioral observations to supplement or occasionally contradict participant self-reports. Click and eye-tracking data were used as qualitative rather than quantitative evidence since they were not uniformly available (see Section \ref{sec:limitations}). 
Theme development drew on latent patterns inferred from behavioral sequences, verbal reasoning, and the researcher's observations. Across three collaborative sessions, the two researchers developed and refined emerging themes. The analysis process (steps 2--5) took roughly 50 hours of work across both researchers. In total, we used 60 codes with seven top-level categories.\footnote{The codebook is available at the OSF Link: \url{https://osf.io/6xqvn/overview?view_only=575d37e431b749df9174e075614d4fcf}}

\subsubsection{Positionality}
The first author has over a decade of experience interacting with children and adults with IDD as a volunteer within the IDD community groups. 
The second author has formal security training and regularly provides volunteer cybersecurity support to at-risk community members, primarily survivors of intimate partner violence.

\subsection{Limitations}
\label{sec:limitations}
This study's small sample (seven participants) and narrow diagnostic range (five primary diagnoses of DS, one IDD, one ASD; see Section \ref{sec:methods:demographics}) only partially covers the broader community \cite{AAIDD, bartoshesky2021intellectual}. Small samples are common in qualitative research with underrepresented groups \cite{Lazar_Hochheiser_Feng_2017}, but limit generalizability; future work should pursue a more diagnostically diverse population.

Five aspects of our data collection may have shaped the behaviors and reports we captured.
\textbf{(1)} Eye-tracking calibration succeeded for only five of seven participants (three with high accuracy) due to vision challenges and attention difficulties, limiting our ability to verify gaze; we therefore used gaze data qualitatively to contextualize attention patterns rather than quantitatively.
\textbf{(2)} We primed participants about the S\&P focus. We used researcher-provided accounts and devices for procedural and ethical reasons (Section \ref{sec:methods:ethical}), unlike S\&P usability studies that use deception or personal devices for realism~\cite{sunshineCryingWolfEmpirical2009}. This may have altered natural interactions, personal stakes, and help-seeking tendencies; to mitigate this, we matched each participant's preferred operating system and customized study tasks (see Section \ref{sec:methods:inclusive}).
\textbf{(3)} We asked participants whether they were already familiar with the notifications encountered; self-reported data is helpful but sometimes unreliable, and given participants' considerable technology experience (see Table~\ref{tab:tech-summary}), retrospective recall may be a limitation here.
\textbf{(4)} We considered only visual notifications on phones, tablets, and laptops; future research should explore accessibility requirements for other modalities, such as auditory or haptic cues.
\textbf{(5)} Researchers offered gentle prompts during moments of avoidance, which may have reduced ecological validity by encouraging persistence with tasks participants might otherwise have abandoned.

Our analysis also does not fully trace evidence to its source within the participant--support person dyad: although prompts targeted participants with IDD (Section~\ref{sec:methods:collection}), support persons sometimes volunteered comments or confirmed routines that we incorporated alongside participant and researcher observations. Since our coding did not consistently attribute evidence by source, findings on interdependence, help-seeking, and household practices may reflect a mix of participant, researcher, and caregiver perspectives. The study also was not designed to examine dyadic or power dynamics, so we cannot assess how support persons influenced participants' stated preferences or hesitancy; future work should attribute evidence by source and examine these dynamics.

\section{Results}

In Section \ref{sec:results:experience}, we review participants' experiences with technology, as well as with S\&P. In Section \ref{sec:results:rq}, we address our central research question about challenges and influential factors across notifications as a whole. For a more detailed breakdown of how participants interacted with each of the notifications they encountered during the study tasks, see Appendix \ref{app_interact}. 
\renewcommand{\arraystretch}{1.5}

\begin{table*}[t]
\caption{Summary of participants’ technology use relevant to the study. \textnormal{This included devices, operating systems, email practices, social media (with only P7 actively using Pinterest), and weather-checking habits. All participants had experience with Apple smartphones, either personally or through family use, and none reported active use of AccuWeather.}}
\Description{Technology use for seven participants (P1--P7), including smartphones, tablets, computers, operating systems, email use, social media, and weather apps. All participants had experience with iPhones or iOS, though one (P6) did not currently use a smartphone. Tablet use (primarily iPads) was reported by four participants (P3--P6). Computer use varied: Windows (P1, P3); Mac (P2, P4); no computer use (P5, P6); unknown systems (P7).
Operating system experience included iOS for all participants, with some also using Windows or macOS. 
Email use ranged from fully independent (P1, P2, P4, P5, P7) to supported by family (P3, P4, P6). Email purposes included work, community coordination, family communication, and social interaction.
Social media use: Instagram (P1); TikTok (P1, P4); Facebook (P1, P2, P6: parents account, P7); Twitch (with support) (P4); and Pinterest (P7); No social media use (P3, P5).
Weather apps: built-in (P1--P3, P5); WeatherBug and news apps (P4); local weather channel apps (P7); none (P6). No participants reported active use of AccuWeather.}
\label{tab:tech-summary}
\centering
\footnotesize
\begin{tabularx}{\linewidth}{l
    >{\raggedright\arraybackslash}p{1.5cm}
    >{\raggedright\arraybackslash}p{1cm}
    >{\raggedright\arraybackslash}p{2.2cm}
    >{\raggedright\arraybackslash}X
    >{\raggedright\arraybackslash}p{3cm}
    >{\raggedright\arraybackslash}X
    >{\raggedright\arraybackslash}X}
\toprule
\textbf{PID} 
& \textbf{Smartphone} 
& \textbf{Tablet} 
& \textbf{Computer} 
& \textbf{OS Experience} 
& \textbf{Email Use} 
& \textbf{Social Media}
& \textbf{Weather Apps} \\ 
\midrule

P1 
& iPhone
& -
& Windows
& iOS;\newline Windows
& Independent\newline Family Community and Work
& Instagram;\newline TikTok;\newline Facebook 
& Built-in phone app \\
\hline
P2 
& iPhone
& -
& Mac (shared)
& iOS;\newline macOS
& Independent\newline Community  
& Facebook 
& Built-in phone app \\
\hline
P3 
& iPhone
& iPad
& Windows (shared)
& iOS;\newline Windows
& Supported by family
& -
& Built-in phone app \\
\hline
P4 
& iPhone
& iPad
& Mac; Prior Chromebook use
& iOS; \newline macOS 
& Independent and supported\newline Social 
& TikTok;\newline Twitch (supported use) 
& WeatherBug;\newline News app \\
\hline
P5 
& iPhone
& iPad
& -
& iOS 
& Independent\newline Family
& - 
& Built-in phone app \\
\hline
P6 
& - 
& iPad
& - 
& iOS 
& Supported by family 
& Views parents’ Facebook 
&-\\
\hline
P7 
& iPhone 
& Unknown
& Unknown
& iOS 
& Independent\newline Community, Social and Work 
& Facebook;\newline TikTok;\newline Pinterest 
& Local weather channel apps \\

\bottomrule
\end{tabularx}

\end{table*}

\subsection{Technology and Security Experience}
\label{sec:results:experience}

\subsubsection{Technology Experience} All seven participants were active technology users with experience with multiple types of devices, shown in Table~\ref{tab:tech-summary}. Tablets were widely used for entertainment, such as YouTube, streaming, and gaming (P3--P7), with some also relying on them for video calls via FaceTime or Zoom (P3, P4). Smartphones were equally common (P1--P5, P7), supporting messaging (P1, P2, P4, P5, P7), social media (P1, P2, P4, P7), work coordination (P2), and mobile banking (P1, P2, P7). Only P7 reported active use of Pinterest. All had experience with Apple smartphones, either their own or their family members’. Four participants routinely used a Windows (P1, P3) or Mac (P2, P4) computer, but it was rarely their primary device. P3 mainly used an iPad but occasionally accessed their mother’s computer for scrapbooking or job training, and P2 shared a Mac with their father. P4 had prior experience with a Chromebook but preferred their iPad.
Technology use often involved family support, such as managing email (P3, P6) or online shopping (P4, P6). At the same time, participants independently used email and calendars to track birthdays, rehearsals, and community events (P1, P2, P4, P5, P7). Checking the weather online was a very popular activity among participants (P1--P5, P7), with none reporting active use of AccuWeather.

\subsubsection{Security Experience}

Participants reported varied approaches to securing their devices and accounts. Device PINs were widely used on smartphones and tablets (P1--P7). Four paired this with biometric methods such as Apple's FaceID or TouchID (P2: ``\textit{I like the face better because I don't remember my passcode,}'' P3, P5, P7).
Account password practices showed differing levels of independence. P1 and P4 relied on password managers built into iCloud and browsers, while P3 and P7 preferred to remember passwords, occasionally writing them down. In contrast, P2 depended on their father to store and update all passwords, and P5 also described interdependence in managing theirs: ``\textit{I ask my mom and dad if I can get help [with passwords and account management], and she could, yeah, a lot, helps out.}'' 

Experiences with account and device lockout were common, and issues were typically resolved with familial support. P1 had been locked out after forgetting passwords for shared family accounts, typically regaining access after calling SP1. P2 described an incident they believed was a Facebook ``\textit{hack,}'' though it was an account lockout; their father helped reset the account. P3 relied on parents to remember updated passwords: ``\textit{He's trying to use old passcode, and it wouldn't work. So then he came up to find Dad}'' (SP3). P4 frequently became locked out of their tablet due to PIN errors and would wait or ask their parents for help. In contrast, P7 reported how they would ``\textit{change my password, if I don't remember it, ... or I call, the-- whoever I need to call and they get me back in}. ''

Participants also described broader security awareness of online scams and fraud. P1 recounted a social engineering attempt on Instagram in which an impostor requested their password, which they refused to share: ``\textit{I was talking to someone that I know with a friend. Whoops, I’m sorry, it was not my friend. There was a friend from college, but it turns out it's not that person. [They messaged me] something like, `what's your password?'}'' P2, P3, and P7 reported receiving scam emails and texts, which they deleted immediately if not from known contacts. For finances, P1 described freezing or locking their debit card if it were misplaced, P7 described having to call customer service to cancel their debit card after seeing suspicious charges, and P3 reviewed banking statements handled by their parents. 

\begin{table*}[t]
\centering\small
\caption{Participant spread across notifications and themes. \textnormal{\textbf{Theme 1}: \textit{Support} = an alternate interpretation traceable to alternative interface elements or elsewhere in the task flow; \textit{Against} = a stable, context-independent understanding, or a personal habit applied identically regardless of a notification's content. \textbf{Theme 2}: \textit{Support} = interpretation or action anchored to an everyday concept, a different specific app/technology, a rule taught by another person, or a stated personal heuristic; \textit{Against} = accurate reasoning grounded in the notification's own content. \textbf{Theme 3}: \textit{Support} = an avoidant, hesitant, diagnostic-exploratory, or support-seeking action driven by not knowing what something meant, or a specific action's outcome; \textit{Against} = confident, decisive action.}}
\label{tab:theme-spread-summary}
\Description{A fourteen-row table listing, for each notification (with 2FA split into its computer, phone-banner, and in-app steps), which participants supported or went against each of the three themes in separate columns, followed by a compact table of six representative examples.}
\setlength{\tabcolsep}{3pt}
\begin{tabular}{l p{2.6cm} p{2.6cm} p{2.6cm} p{2.6cm} p{2.6cm}p{2.6cm}}
\toprule
& \multicolumn{2}{c}{\textbf{Theme 1: Task \& context}} & \multicolumn{2}{c}{\textbf{Theme 2: Everyday anchors}} & \multicolumn{2}{c}{\textbf{Theme 3: Uncertainty}} \\
\cmidrule(lr){2-3} \cmidrule(lr){4-5} \cmidrule(lr){6-7}
\textbf{Notif.} & \textbf{Support} & \textbf{Against} & \textbf{Support} & \textbf{Against} & \textbf{Support} & \textbf{Against} \\
\midrule
2FA-C & & P6 & P6 & P4 & P1, P2, P3, P4, P6 & \\
2FA-B & P2, P4 & P1, P3 & P1, P3, P7 & P4, P6 & P2, P7 & P6 \\
2FA-I & P2 & P1, P3, P7 & P1, P7 & P3 & P2, P6 & P1, P3, P4, P7 \\
SP  & P6, P7 & P3, P5 & P1, P2, P3, P4, P5, P6 & & P4, P5 & P1, P2, P3, P6, P7 \\
CC  & P1, P2, P4 & P6 & P3, P5, P6 & P2 & P2, P3, P4, P7 & P1, P5 \\
MW  & & P1, P2, P3 & P1, P2, P3 & P4, P6 & P2, P3, P6, P7 & P1, P4, P5 \\
DD  & P1 & P2, P7 & P2, P4, P5 & P3 & P1, P2, P3, P4, P5, P6, P7 & \\
HT  & P2, P4, P5 & P1, P7 & P1, P5, P6 & P7 & P4, P6 & P1, P2, P3, P5, P7 \\
TP  & P1, P2, P4, P5, P7 & & P1, P2 & P3, P7 & P2, P4, P6, P7 & P1, P5 \\
LP  & P4 & P1, P2, P3, P5 & P1, P2, P5 & & P1, P3, P6 & P2, P4 \\
NP  & P4 & P1, P2, P3, P5, P6, P7 & P2, P3, P4, P5, P6 & P7 & P3 & P1, P4, P5, P6, P7 \\
NSI & P2, P4 & P1, P3, P7 & P2, P5 & P1, P3, P4, P7 & P2, P4, P5, P6 & P1, P3, P7 \\
SRI & P2, P4, P5, P6, P7 & P1 & P5 & P1, P2 & & P1, P2, P4 \\
MUI & P3, P4, P6, P7 & P1 & P1, P6 & & P3, P7 & P1, P6 \\
\bottomrule
\end{tabular}

\vspace{4pt}
\begin{tabular}{p{1.6cm} p{7.5cm} p{8.0cm}}
\toprule
 & \textbf{Support example} & \textbf{Against example} \\
\midrule
Theme 1 & P4 (LP): read ``tracking'' as being about the weather rather than her own location, holding to that interpretation even after two direct redirections. & P1 (MW, also HT): confirmed he always clicks out of a warning without reading it, a fixed habit applied identically regardless of a notification's content (``\textit{Do you ever click that button?}'' ``\textit{No.}'' ``\textit{You just click out of it like you did today.}'' ``\textit{Yeah, exactly.}''). \\
Theme 2 & P6 (CC, also P3, P5): ``cookies'' read as food (``\textit{chocolate chip}''), opposed to the technical meaning. & P3 (NSI): matched the login email's city and state to where he actually lives (``\textit{this one I am in... this one I am not in}''), correctly identifying the fraudulent email through notification content. \\
Theme 3 & P2 (2FA-C): paused seven seconds before a nervous ``\textit{Kayyy,}'' then asked, ``\textit{Can I do this one?}'' before acting. & P7 (HT): correctly distinguished HT from MW using each notification's text (``\textit{both of them are dangerous, but this one is not private}''), then went straight to ``Back to Safety'' without hesitation. \\
\bottomrule
\end{tabular}
\end{table*}

\subsection{Experiences, Challenges, and Interaction Decisions with S\&P Notifications}

\label{sec:results:rq}

In this section, we answer \textit{RQ: How do adults with IDD experience S\&P notifications, what challenges do they face, and what factors influence their understanding and decision-making?}
Participants encountered a range of S\&P notifications embedded in everyday tasks, providing insight into their decision-making processes. Challenges arose less from isolated design features than from the ways notification content, context, and task goals combined to shape understanding and choice. Three themes capture these challenges and the factors influencing decision-making: (1) notification was interpreted through the ongoing task and interface context; (2) everyday concepts provided anchors for interpreting unfamiliar terminology; and (3) exploratory and avoidant actions emerged when participants were unsure of the action outcomes. We describe each theme below, with Table~\ref{tab:theme-spread-summary} summarizing which participant spread per notification across every theme.

\subsubsection{Theme 1: Notification Interpretation Was Shaped by the Ongoing Task and Interface Context}
\label{results:theme1}
Participants repeatedly interpreted notification content and interaction options through the task they were performing (see Section~\ref{sec:methods:taskdesign}). Interpretation was anchored through the surrounding screen or the task flow (P2--P7). 
On the status bar, participants read the screen-recording indicator (SRI) through the overlapping clock face (P2, P4--P7): P2 explained the red color meant `\textit{`the time is blocked, it can stop that,''} and P6 guessed, `\textit{`maybe it's three... it's around the clock, right?''} The microphone-in-use indicator (MUI) was read similarly through the neighboring Wi-Fi or battery icon (P3, P4, P6, P7); P6 guessed \textit{``for charger,''} explaining that orange commonly signals a low battery on their own devices, borrowing meaning from whichever familiar icon was closest. 

Meaning also carried over through the task flow rather than the notification itself (P1, P2, P4). For example, P1 explained the dangerous download (DD) warning as \textit{``a blocked site,''} because he clicked a button ``Visit Site.''; P2 and P4 evaluated the login alert (NSI) as spam rather than identity verification after completing spam detection in the previous task (P2: \textit{``Yeah, this is, um, spam.''}). The location permission (LP) similarly shaped how three participants (P1, P2, P4) read the cross-app tracking permission (TP) notification moments later (\textit{ex.,} Table~\ref{tab:theme-spread-summary}); and the cookie-consent (CC) banner was treated as a login gate rather than a privacy choice (P1, P2, P4) with P2 chosing Allow because the banner said cookies were needed \textit{``to use our site.''}

Prior work has shown that users often follow the apparent task flow of an interface rather than treating notifications as independent decisions, especially in cookie-consent and dark-pattern scenarios where buttons like Accept All appear as the only path forward \cite{berens2024cookie, habib2022okay, utz2019UninformedConsent}. Our findings extend this work by showing that flow-based interpretation also occurred across diverse security notifications, even without dark patterns. 

\subsubsection{Theme 2: Everyday Concepts Provided Anchors for Understanding and Action}
\label{results:theme2}

When S\&P notification semantics or semiotics resembled something participants already understood, those familiar associations framed their interpretation of the notifications. For example, cookies (CC) were read through the food concept regardless of the banner's actual content (\textit{ex.,} Table~\ref{tab:theme-spread-summary}). 
The cross-app tracking permission request (TP) and notification requests (NP) were interpreted through everyday forms of social sharing (P1, P2, P5) with P2 grounding terminology,  \textit{``across other companies,''} in their family's real Spotify-sharing practice, extending the idea to shared devices and even \textit{``your next-door neighbor.''} 

Heuristics learned from family or work shaped how participants assessed S\&P notifications (P1--P7). P1 and P3 were taught spam (SP) equates to ``\textit{scam,}'' while P4 associated spam with strangers or ``people trying to hack into, like, a Facebook account.'' Color carried similar everyday weight \cite{edworthy1996WarningDesign, wogalter2002ResearchbasedGuidelines}: P3, P4, P5, and P6 all flagged the red spam banner (SP-R) as most dangerous on sight, with P3 audibly startled (\textit{``Ooh! Th-th-th-this is is v-very important''}); and P7 read yellow as \textit{``be careful.''} Not everyone leaned on color. P1 consistently said the color made no difference to him and looked to the sender or email content instead. Interestingly, sound anchored meaning for P6 when notification (NP) was reframed as the sound his own tablet makes (\textit{``when it goes ding!''} SP6).

Existing accessibility and usable security guidance emphasizes avoiding jargon and simplifying wording \cite{wcag22, chang2021towards, still2017human, garfinkel2014usable, sp2005usability}. Our findings extend this work by showing that participants grounded unfamiliar language in experience-based anchors. These anchors helped participants reason about unfamiliar security concepts, even when the analogy was misleading (\textit{e.g.,} TP is not related to your personal location; interacting with DD will not ``block you out, the entire computer... open up different viruses'' (P2), \textit{etc.}). 

\subsubsection{Theme 3: Uncertainty About Meaning or Outcome Prompted Hesitation, Diagnostic Exploration, Avoidance, and Support-Seeking}

Participants avoided, hesitated, explored, or directly requested help when they did not know what a notification meant (\textit{comprehension}) or what an action would lead to (\textit{outcome}).  Comprehension gaps most often showed up as visible hesitation before acting (\textit{e.g.,} pauses, unsettled gazes, or brief periods hovering over an element) across nearly every participant (P1--P4, P6, P7): on the MW, P3 hovered over the Details link, then pulled back, calling it \textit{``the scary thing.''}, (\textit{ex.,} Table~\ref{tab:theme-spread-summary}). Outcome uncertainty -- not knowing what an action would do -- appeared across three participants (P2, P4, P5): on the DD, P2 let it disappear rather than click it, explaining they \textit{``did not know what clicking on it would do''}; and P5, on the NSI, declined to click a link to \textit{``secure your account'}' they had correctly identified as clickable. Diagnostic exploration -- clicking to test action-outcome -- appeared across two participants (P1, P6). For example, P6 clicked non-functional elements on the 2FA-I, MW, and the NSI before resolving each. 

This was not universal. Quick, confident interactions split into three patterns: (1) dismissive of notifications (P1, P5): P1 summed up their approach as \textit{``I just don't care''}; P5 clicked through the MW just as fast, with a flat, disengaged reaction; (2) failed to notice notifications (P1--P7): P6 did not fixate on DD reporting,\textit{``[it's] too small... it blends in too much,''} and no one reported noticing SRI or MUI; and (3) readily understood the S\&P notifications (P3, P4, P7): P4 named the MW's danger immediately from its title and red coloring alone; and P7 correctly distinguished the HT from the MW using each notification's own text (\textit{``this one is not private''}). 

Participants engaged interdependently with family, friends, and trusted others to seek or provide guidance about \spn notifications (P2--P4, P6). In-session, this took the form of direct requests (\textit{e.g.,} P2 asked \textit{``Can I do this one?''} before 2FA-C) mirroring routines described outside the study. SP3 explained \textit{``[P3] will come and ask one of us... before opening [unknown emails],''} and SP4 described a reciprocal relationship (\textit{``Sometimes [P4] shows me''}). P3 generalized this as normal practice, saying they would \textit{``either do it on their own or ask someone.''} Interdependence did not guarantee assistance(SP4: \textit{``I don't even know''} on MUI), but it was a consistent strategy.

Avoidance and uncertainty are documented security behaviors \cite{liang2010understanding}, but our findings show how these behaviors surfaced for participants with IDD, with notifications often providing limited support for resolving uncertainty.

\section{Discussion}
While designing and conducting the study, additional tensions and accessibility challenges within \spn became clear, beyond our three themes.

First, participants often expressed a desire for safety (\textit{e.g.,} ``\textit{I'm trying not to get hacked}''), but this often produced cautious inaction or exploratory clicking rather than the intended secure action. Most avoided perceived unsafe actions by dismissing or ignoring a notification, with one exception: P6, who clicked through to a simulated phishing link. However, this caution is not uniformly protective. For new sign-in alerts (NSI), where the secure response requires acting rather than avoiding, the same hesitation could leave participants exposed, echoing prior work showing inaction toward warnings can itself \textit{undermine} security~\cite{bravo-lilloHarderIgnoreRevisiting2014}. That work attributes inaction to habituation or warning fatigue \cite{bravo-lilloBridgingGapComputer2011,reederExperienceSamplingStudy2018}. Our findings suggest a related but distinct pathway for participants with IDD. Avoidance may be rooted not only in fatigue, but in genuine uncertainty about a notification's meaning or an action's outcome. This distinction may call for different interventions: novelty or interruption for fatigue, and decision support for uncertainty.

Our findings expose potential limitations in how existing accessibility standards, such as the Web Content Accessibility Guidelines (WCAG)~\cite{wcag22}, address S\&P notifications. WCAG emphasizes that interfaces should be perceivable, operable, and understandable, but we observed breakdowns in all three. Subtle recording indicators were imperceptible, hyperlinks with actions like ``Report this'' were overlooked (operability), and abstract terms such as ``verification'' compromised understandability. 
Because notifications are brief, secondary, and time-sensitive, they may require tailored accessibility principles distinct from the large-scale systems WCAG was designed around. 
As prior accessible web content design has shown \cite{Karreman_Van_Der_Geest_Buursink_2007}, linguistic simplicity alone does not guarantee comprehension. Our findings point to the value of situated semantic design, where meaning stays consistent across shifting visual and task contexts. We also observed patterns consistent with existing IDD accessibility guidance -- large, clearly labeled elements and minimal interactive choices \cite{chang2021towards} -- but participants still struggled when notifications failed to separate from the ongoing task. Although animations are often recommended~\cite{sundeepa2021effectiveness}, the abstract 2FA animation was frequently ignored, suggesting animations may need to be realistic and action-oriented. Future research could investigate whether audio or haptic options would be beneficial as well. 

S\&P notifications, though considered together here, often have competing design goals. Most security notifications nudged users toward protective choices, whereas many privacy notifications serve a business interest in permissive settings by exploiting cognitive biases or misconceptions~\cite{nouwens2020DarkPatterns}. Users often accept cookie banners believing this is required for the site to function~\cite{utz2019UninformedConsent}; two participants feared denying cookies would log them out or break the site, citing notification phrasing that reinforced the misread. Yet even in these privacy contexts, accessibility and commercial incentives are not entirely irreconcilable. As accessibility becomes a regulatory requirement (\textit{e.g.,} WCAG 2.2, EU Digital Services Act, Section 508), compliance provides a rationale for clearer designs, and strengthening consumer trust and retention may offset short-term data losses. While accessibility cannot resolve the incentives for dark patterns~\cite{brignullDP2023}, it can limit their most harmful forms and set a normative baseline for transparency and accountability.

\begin{table*}[t]
\centering\small
\caption{Mapping from the three findings themes (Section~\ref{sec:results:rq}) to the three design implications (Section~\ref{sec:designimp}), with a representative behavioral example for each theme--implication connection.}
\label{tab:theme-design-mapping}
\Description{A table with three row-groups, one per design implication, each listing the theme(s) that motivate it alongside a concrete participant example.}
\begin{tabular}{p{3.3cm} p{2.6cm} p{9.7cm}}
\toprule
\textbf{Design Implication} & \textbf{Theme} & \textbf{Representative Example} \\
\midrule
\multirow{2}{3.3cm}{\textbf{\#1:} Address Context-Dependent Language Misunderstandings}
 & \multirow{2}{2.6cm}{Theme 1: Task \& Context}
   & P4 read the location-permission prompt (LP) through the weather app, explaining ``tracking'' as being about the weather -- and held that reading even after the researcher tried twice to redirect her. \\
\cmidrule(lr){2-3}
 & \multirow{2}{2.6cm}{Theme 2: Everyday Anchors}
 & P1 mapped the word ``spam'' directly onto ``scam,'' a term he had learned through formal job training, rather than the notification's own framing. \\
\midrule

\multirow{3}{3.3cm}{\textbf{\#2:} Make Action-Outcome Connections Transparent}
 & Theme 1: Task \& Context
   & On a 2FA, P2 hesitated because she ``\textit{didn't know if it was [the computer] or [the phone]}'' -- task-context confusion about which device's action would actually register. \\
\cmidrule(lr){2-3}
 & \multirow{2}{2.6cm}{Theme 3: Uncertainty}
   & P2 let the dangerous download warning disappear rather than click it, explaining she ``\textit{did not know what clicking on it would do.}'' \\
 & & P5 recognized the ``secure my account'' link on the NSI as clickable, but declined to click it specifically to avoid an unknown outcome. \\
\midrule

\multirow{4}{3.3cm}{\textbf{\#3:} Enable Interdependent Decision-Making}
 & \multirow{2}{2.6cm}{Past Technology Experiences}
   & P2 described receiving messages from strangers over Facebook Messenger asking for money (\textit{``Yeah, my money and information''}) and requested a parent-facing safeguard to identify spam: \textit{``a button that [users] can push on, or their parents can do that.''} \\
\cmidrule(lr){2-3}
 & \multirow{2}{2.6cm}{Theme 2: Everyday Anchors}
   & SP3 confirmed that P3 has a standing household practice of checking with a support person before opening an uncertain email: ``\textit{He'll come and ask one of us... before opening it.}'' \\
\cmidrule(lr){2-3}
 & \multirow{2}{2.6cm}{Theme 3: Uncertainty}
   & P3 asked the researcher outright ``\textit{which one should I pick?}'' on the cookie-consent banner rather than deciding independently. \\
 & & P2 asked ``\textit{Can I do this one?}'' before attempting the two-step-verification prompt, seeking reassurance before acting. \\
\bottomrule

\end{tabular}
\end{table*}

\subsection{Design Implications}
\label{sec:designimp}
By examining participants’ interactions with a range of S\&P notifications (Appendix~\ref{app_design} and \ref{app_interact}) and identifying challenges and factors that influenced their engagement (Section~\ref{sec:results:rq}), we derived three design implications for more inclusive security and privacy notifications (Table~\ref{tab:theme-design-mapping}).

\subsubsection{\textbf{Implication \#1: Address Context-Dependent Language Misunderstandings Beyond Jargon Simplification}}
\textit{Security and privacy notifications should go beyond avoiding technical jargon to account for how context shapes meaning.} Participants’ misunderstandings, such as equating ``cookies'' with food or interpreting ``deny'' as ``log out,'' did not arise purely from vocabulary deficits, but from contextual blending between task, prior experience, and interface, described in Themes~1 and~2. Existing guidance emphasizes simplifying language \cite{Bauer2013, chang2021towards, wcag22}, but our findings suggest this alone may be insufficient. Designers should therefore treat clarity as context calibration by explicitly identifying what is being acted upon (``your phone's location,'' ``this laptop'') and using progressive disclosure when additional clarification is needed. Visual and multimodal cues could further reinforce this meaning by distinguishing \spn notifications from surrounding interface content (\textit{e.g.,} a temporary boundary highlight or audio cue could distinguish notifications from surrounding interface content). By communicating meaning through language, visuals, and interaction, notifications may reduce context-driven misunderstandings that simplified language alone cannot resolve.
\subsubsection{\textbf{Implication \#2: Make Action-Outcome Connections Transparent}}

\textit{Notifications should clearly connect the security actions a user can take to the actions' outcomes.} Participants' uncertainty revealed that action verbs alone (\textit{e.g.,} ``Secure your account'' or ``Learn more'') do not foster confidence when outcomes are opaque. This need emerged across Themes~1 and~3 (Table~\ref{tab:theme-design-mapping}), where uncertainty about action outcomes led participants to hesitate, avoid acting, or seek help. Existing guidance recommends concise, command-style options \cite{Bauer2013}, while prior work on notification design for cognitively diverse users emphasizes timing, modality, and situational framing to support comprehension and trust \cite{eskandar2022designing, rehm2024notification, jamieson2020technology}; multimodal cues have also been shown to reinforce comprehension \cite{yu2022effects, fujisawa2011effect, eskandar2022designing}. Our findings extend this work by suggesting that accessible S\&P notifications should communicate not only the security actions available to users, but also the consequences of those actions before they are taken. Designers could achieve this by communicating the resulting system state without revealing sensitive information or requiring lengthy explanations. For example, a brief preview of the next screen or an optional guided walkthrough could help users anticipate the outcome of an action before committing to it.

\subsubsection{\textbf{Implication \#3: Enable Interdependent Decision-Making}}
\label{section:dis:di3}

\textit{S\&P mechanisms should be designed to allow users to rely on trusted others and to alleviate time and uncertainty pressures.}
Adoption of S\&P features and responses to security concerns are deeply social and interpersonal experiences for many users ~\cite{redmilesShouldWorryCrossCultural2019, redmilesThinkTheyreTrying2016, mendel2022ProactiveSupporta, dasRoleSocialInfluence2015}. 
This became clear during our study as well. Participants already had established ways of handling uncertainty by turning to a trusted person, whether through a standing household practice (Theme~2 and Section~\ref{sec:results:experience}) or a request for help in the moment (Theme~3).
Participants often drew on heuristics learned from caregivers or trusted individuals (\textit{e.g.,} spam = scam, never open marketing emails), paused to ask what their support person would do, or sought explicit affirmation before performing an action. This is consistent with prior work on how adults with IDD make S\&P decisions with caregivers \cite{sundeepa2021effectiveness, venkatasubramanian2021exploring, chang2021towards, chalghoumi2019information, rocheleau2021caregivers}
Crucially, in our study, participants flipped this dynamic, routinely helping their parents or peers with tech-related tasks (P1--P4). We view this phenomenon through the frame of \textit{interdependence for assistive technology (AT)} introduced by ~\citeauthor{bennett2018InterdependenceFrame}, which posits that ``independence'' is \textit{not} the only form of autonomy for users with disabilities, and that giving/seeking support are important expressions of agency. Participants often described working with others when navigating S\&P decisions, particularly around authentication and consent (P1--P6). For example, when asked what they would do if locked out of an account, P1 replied, \textit{`call for assistance... family, of course... we all need to rely on someone.''} One participant (P7) more often described resolving authentication challenges independently (\textit{e.g.,} \textit{``I call the [customer service]-- whoever I need to call and they get me back in.''}), suggesting that interdependent security practices were common, but not universal, among our participants.

We argue that S\&P notifications can and should flexibly support interdependence when users are faced with decisions (\textit{e.g.,} whether a button is safe to click, which privacy settings make sense for them). This does not involve taking away control from the user, but rather reducing friction, confusion, and stigma if and when users give/seek support.
For example, time-sensitive notifications, such as 2FA, should clearly signal that exiting the flow is safe and the user can try again later after they have received support. Notification text can reduce stigma by acknowledging confusing terms and avoiding language that makes users feel incompetent. Designing for such flexible stewardship (\textit{e.g.,} asking the primary user to designate a collaborator during application setup, user-initiated controls, time-based access, \textit{etc.}) could balance independence with the reality of interdependent security practices. However, care must be taken in the design of such a multi-user system to avoid creating new avenues for technology abuse~\cite{naseri2019accessileaks,jang2014a11y, Patrus2021IllegibleInjury}.

\section{Conclusion}

We conducted a formative user study with seven adults with IDD to understand how they perceive and interact with common security and privacy notifications in email, social media, and mobile apps. Thematic analysis revealed that participants interpreted notifications through the task and interface context, mapped both technical and non-technical terms to everyday concepts, and uncertainty about meaning and consequences drove hesitation, avoidance, and support-seeking. We present three design recommendations: (1) address context-dependent language misunderstandings beyond jargon simplification; (2) make action-outcome connections transparent; and (3) enable interdependent decision-making. We call on researchers and developers to adopt these principles to make security and privacy notifications accessible for adults with IDD.

\begin{acks}
 This material is based upon work supported by the \textit{National Science Foundation Graduate Research Fellowship Program} under Grant No. DGE-2137424, the National Science Foundation award \#2339679 and the Baldwin Wisconsin Idea Endowment, and the Sheldon B. \& Marianne S. Lubar Professorship awarded to Bilge Mutlu. Any opinions, findings, conclusions, or recommendations expressed in this material are those of the author(s) and do not necessarily reflect the views of the National Science Foundation. We would like to thank the volunteer families who participated in our study, and the anonymous reviewers who provided thoughtful feedback on our paper.
\end{acks}

\bibliographystyle{ACM-Reference-Format}
\bibliography{bib}

\appendix
\onecolumn

\clearpage
\section{Notifications Shown During Study Session} \label{app_notifs}

\newcommand{\myincludegraphics}[1]{ \includegraphics[width=\linewidth, height=6cm, keepaspectratio]{#1}}

\newcommand{\subfigsize}{0.32\linewidth}
\newcommand{\leftwidth}{0.365\linewidth}
\newcommand{\midwidth}{0.365\linewidth}
\newcommand{\rightwidth}{0.26\linewidth}

\newcommand{\subfigcaption}[2]{\caption{\textbf{#1} ~~ --- ~~ #2}}

\newcommand{\twoFAfigures}{
    \begin{figure}[h]
        \centering
        \Description{Screenshots of three variants of Google 2-factor authentication prompts, described below.}
        \caption{Two-factor Authentication Notifications}
    \begin{minipage}[c]{0.62\linewidth}
        \centering
        \begin{subfigure}[t]{\linewidth}
            \centering
            \fbox{\includegraphics[width=\linewidth, height=6cm, keepaspectratio]{figs/subfigs/2fac.png}}
            \Description{Screenshot of a Google UI with two columns. On the left, note text saying ``2-Step Verification. To help keep your account safe, Google wants to make sure it's really you trying to sign in.'' On the right, there is an illustration of the Gmail app opening on a phone. Below the image there is text: "Check your Galaxy S20+5G. Google sent a notification to your Galaxy S20+5G. Tap YES on the notification to verify it's you" and a checkbox next to text "Don't ask again on this device."}
            \caption{\textbf{2FA-C} ~~ --- ~~ Computer 2FA interruption while logging into Gmail}
            \label{fig:2fa4:c}
        \end{subfigure}
    \end{minipage}
    \begin{minipage}[c]{0.32\linewidth}
        \centering
        \begin{subfigure}[t]{\linewidth}
            \centering
            \fbox{\includegraphics[width=0.8\linewidth, height=4cm, keepaspectratio]{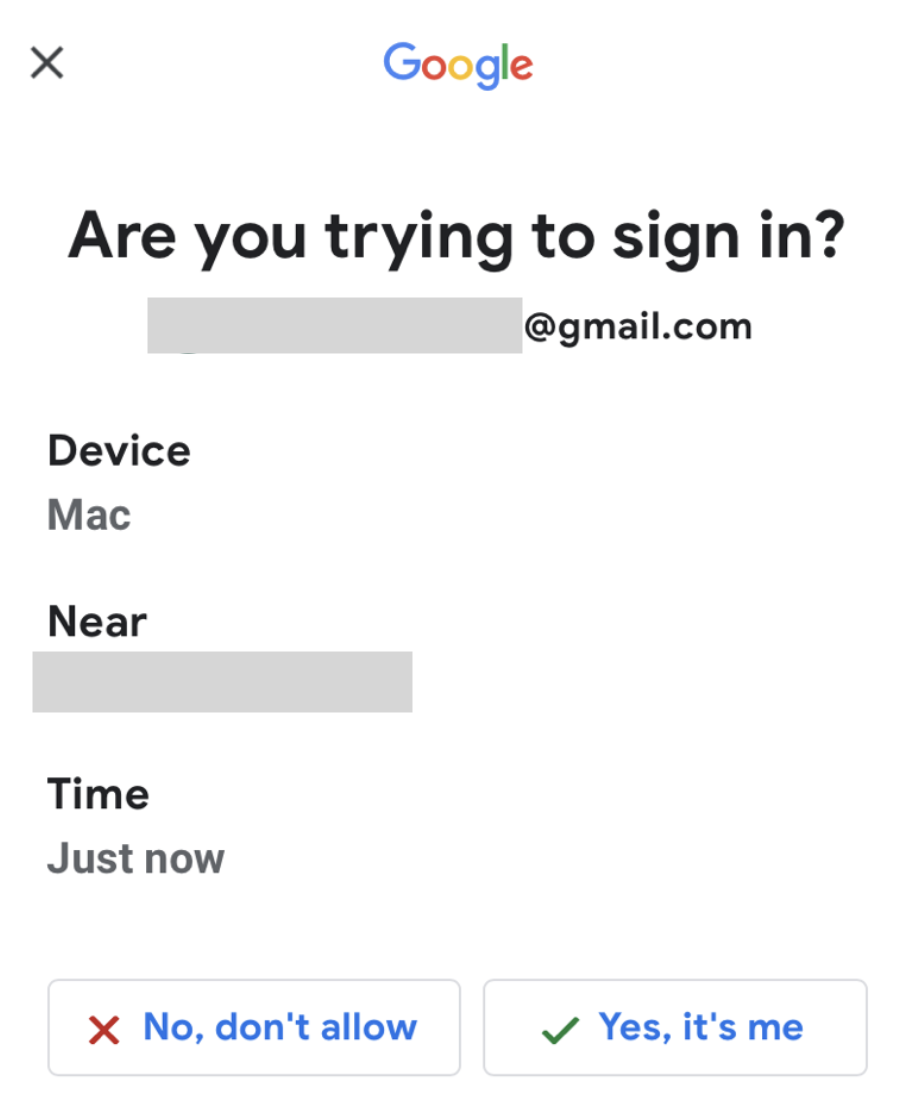}}
            \Description{Screenshot of a Google mobile UI with text and buttons. From top to bottom: Text "Are you trying to sign in?" with the email account listed. Then text: "Device: Mac, Near: [redacted location], Time: Just now". Below this text, two side-by-side buttons, the left-most saying "No, don't allow" with a red X icon and, on the right, "Yes it's me" with a green checkmark.}
            \caption{\textbf{2FA-I} ~~ --- ~~ In-app approval}
            \label{fig:2fa4:i}
        \end{subfigure}
        \\ \vspace{.5em}
        \begin{subfigure}[t]{\linewidth}
            \centering
            \fbox{\includegraphics[width=0.8\linewidth, keepaspectratio]{figs/subfigs/2fa-banner.png}}
            \Description{Screenshot of a push notification banner from Gmail saying "Time sensitive. Confirm whether it's you trying to sign in [redacted email address]" with Gmail's logo (color envelope) on the left.}
            \caption{\textbf{2FA-B} ~~ --- ~~ Phone push banner}
            \label{fig:2fa4:b}
        \end{subfigure} 
    \end{minipage}
    \label{fig:2fa4}
    \end{figure}
}

\newcommand{\chromefigures}{
    \begin{figure}[]
        \centering
        \Description{Screenshots of three variants of browser warnings originating from Chrome, described below.}
        \caption{Chrome Browser Warnings}
        \begin{subfigure}[t]{0.49\linewidth}
            \centering
            \includegraphics[width=\linewidth, height=4.8cm, keepaspectratio]{figs/subfigs/malicioussite.png}
            \Description{Screenshot of bright red Chrome browser warning taking up the whole screen. At the top, there is a white octagon icon with a red X in the middle with large white font "Dangerous site" below it. The paragraph below says:  "Attackers might be trying to steal your information from expired.badssl.com (for example, passwords, messages, or credit cards)." with a hyperlink stating "Learn more about this warning." Below this there are two oval buttons: "Advanced" which is the same color as the background and unemphasized, and "Back to Safety," which is blue and visually prominent.}
            \caption{\textbf{MW} ~~ --- ~~ Malicious website blocked}
            \label{fig:chrome4:mw}
        \end{subfigure}\hfill
        \begin{subfigure}[t]{0.5\linewidth}
            \centering
            \includegraphics[width=\linewidth, height=4.8cm, keepaspectratio]{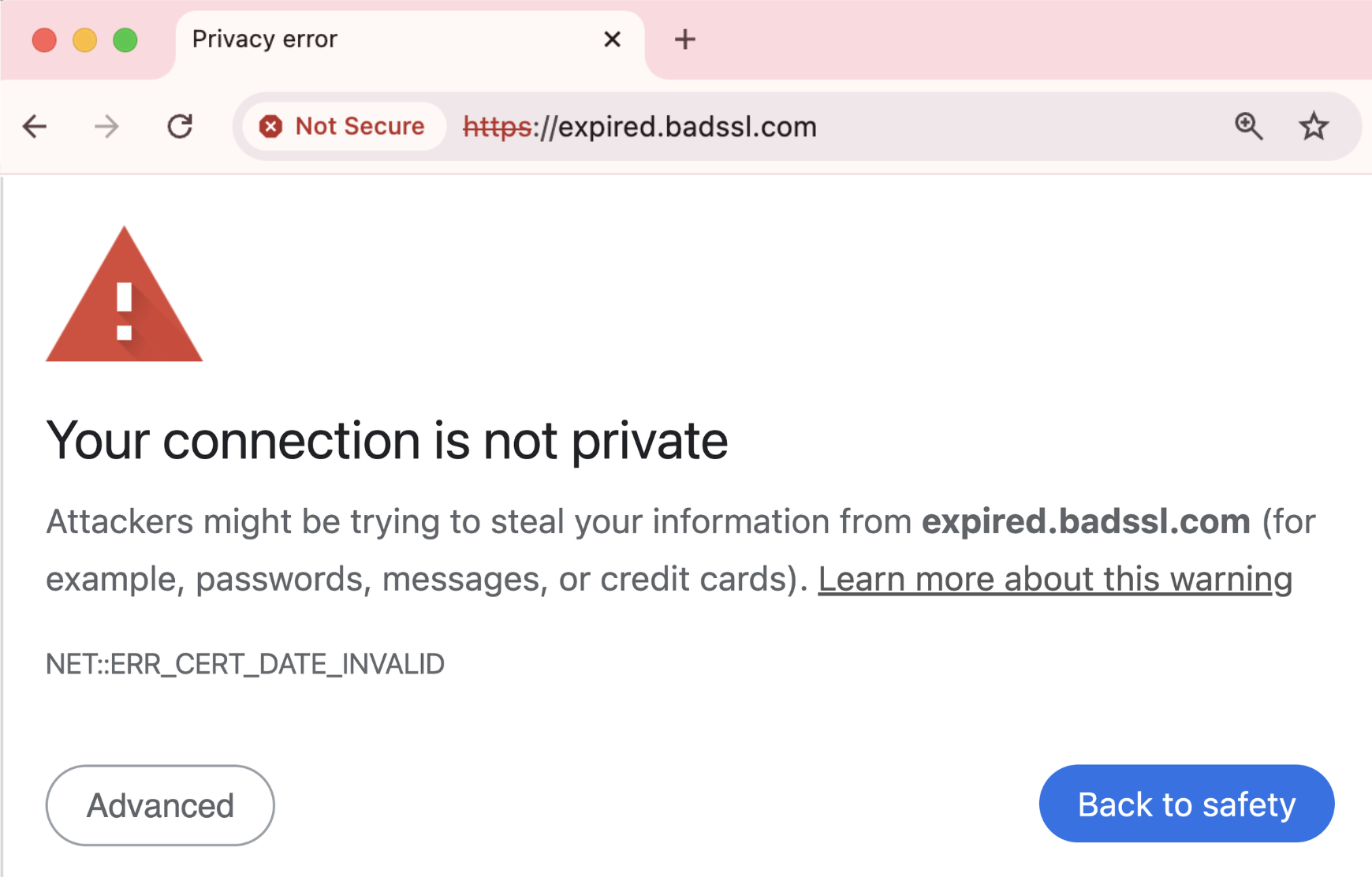}
            \Description{Screenshot of Chrome browser warning taking up the whole screen with a white background. At the top, there is a red triangle icon with a white exclamation point in the middle with large red font "Your connection is not private" below it. The paragraph below says:  "Attackers on the site you tried visiting might install harmful software that steals or deletes things like your passwords, photos, messages, or credit card numbers. Chrome strongly recommends going back to safety." with a hyperlink stating "Learn more about this warning." Below this there are two oval buttons: "Details" which is the same color as the background and unemphasized, and "Back to Safety" which is white and visually prominent.}
            \caption{\textbf{HT} ~~ --- ~~ Invalid HTTPS certificate}
            \label{fig:chrome4:ht}
        \end{subfigure}
        \\ \vspace{.5em}
        \begin{subfigure}[t]{0.48\linewidth}
            \centering
            \includegraphics[width=\linewidth, keepaspectratio]{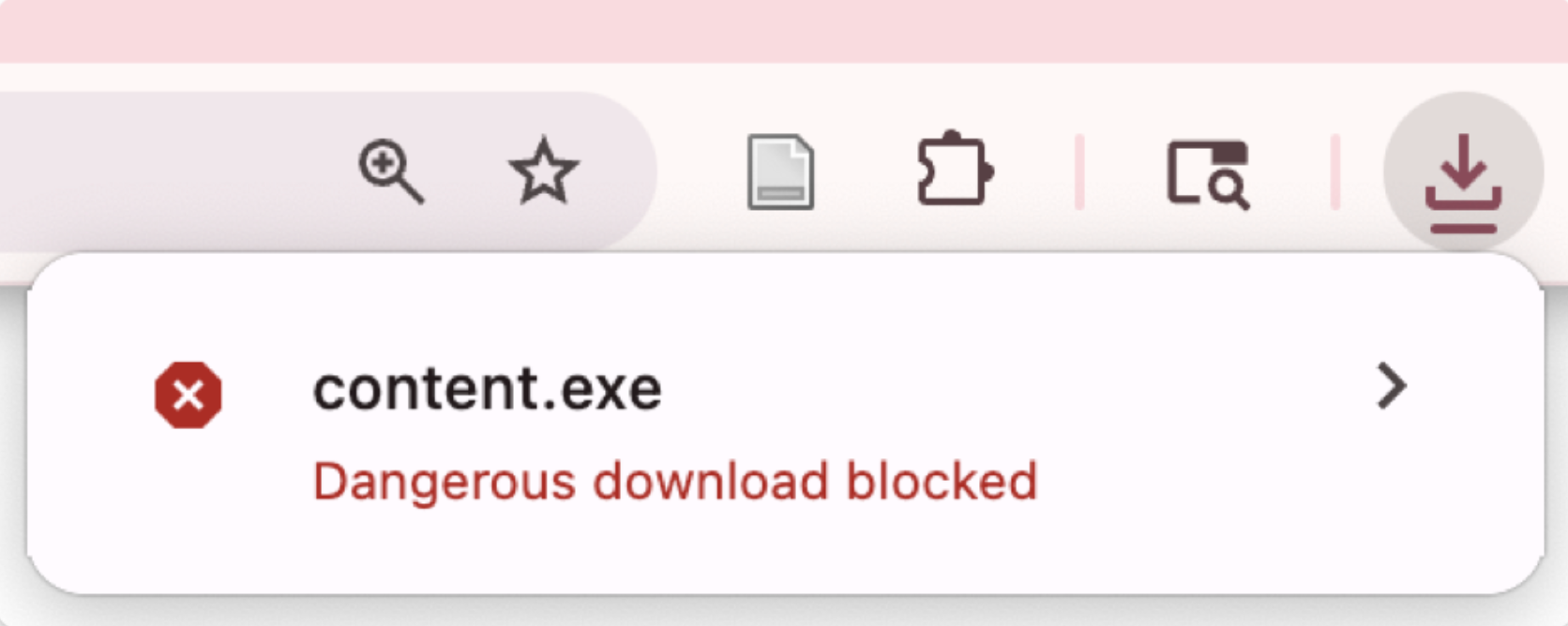}
            \Description{Screenshot of a pop-up warning that appears in the top right of the Chrome browser near the downloads tab. It is a white rectangle with an icon on the left (red octagon icon with a white X) with the file name "content.exe" "Dangerous download blocked" in the middle. On the right, there is an arrow that expands the pop-up with more informational links.}
            \caption{\textbf{DD} ~~ --- ~~ Dangerous download blocked}
            \label{fig:chrome4:dd}
        \end{subfigure} 
    \label{fig:chrome4}
    \end{figure}
}

\newcommand{\iospermissionfigures}{
    \begin{figure}[]
        \centering
        \Description{Screenshots of three standard iOS system prompts, described below.}
        \caption{iOS Permission Requests}
        \begin{subfigure}[t]{0.32\linewidth}
            \centering
            \includegraphics[height=5.4cm, width=\linewidth, keepaspectratio]{figs/subfigs/crossapptracking.png}
            \Description{Screenshot of an iOS notification pop-up requesting permission to ``Allow AccuWeather to track your activity across other companies and websites?'' Underneath it says ``This permission does not increase the number of ads you see, but personalizes the ones you do.'' Underneath it has two buttons, The top one says ``Ask app not to track'' and the bottom one says ``Allow.'' The notification appears in the middle of the iPhone screen.}
            \caption{\textbf{TP} ~~ --- ~~ Cross-app tracking permission}
            \label{fig:ios:tp}
        \end{subfigure}\hfill
        \begin{subfigure}[t]{0.32\linewidth}
            \centering
            \includegraphics[height=6.3cm, keepaspectratio]{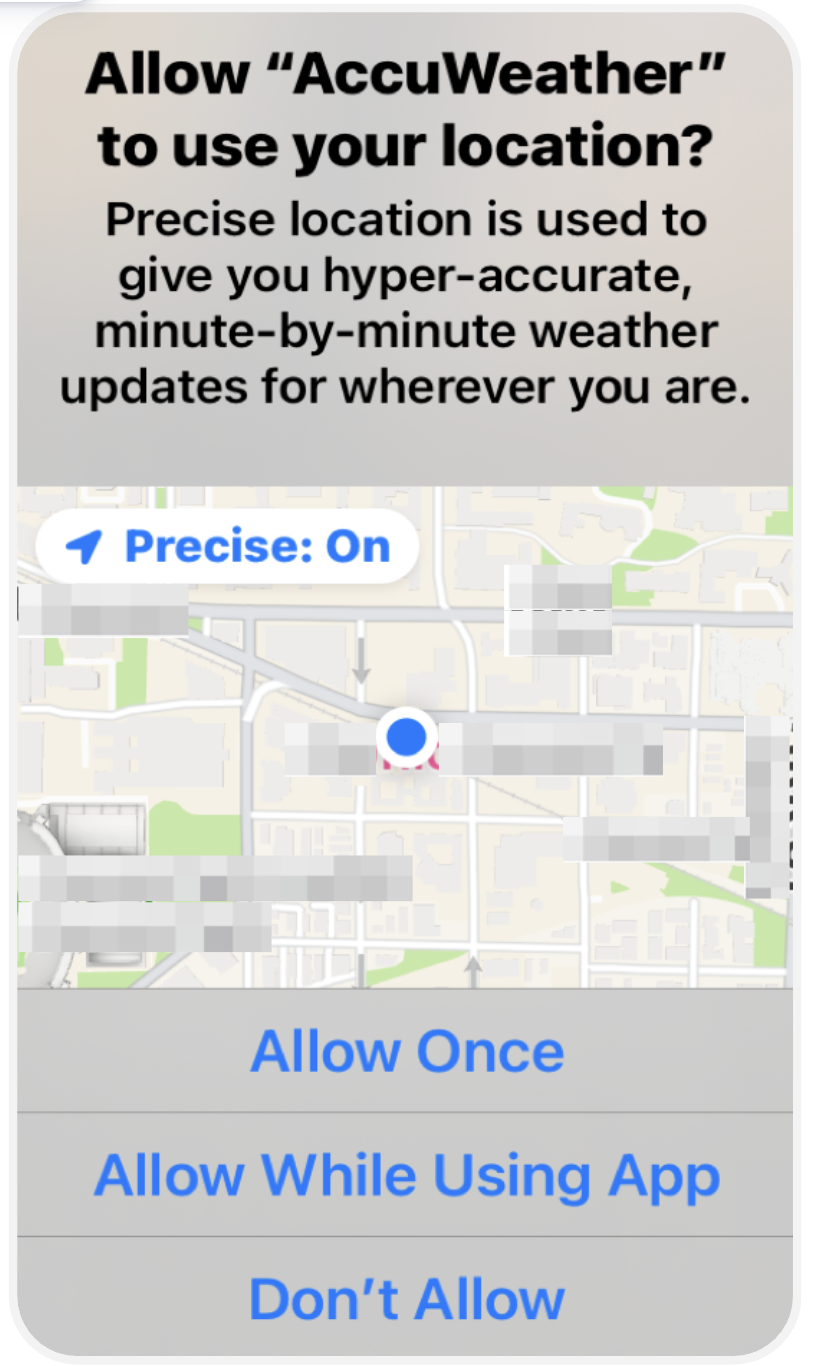}
            \Description{Screenshot of an iOS pop-up asking for location access: "Allow "AccuWeather" to use your location?" The message states that precise location is used to give hyper-accurate, minute-by-minute weather updates wherever the user is. A miniature map is displayed beneath the text, showing a blue dot indicating the user's location. Three buttons are shown at the bottom, on top of each other: "Allow Once," "Allow While Using App," and "Don’t Allow," with the last option placed at the bottom. The notification appears centered on the iPhone screen.}
            \caption{\textbf{LP} ~~ --- ~~ Location permission}
            \label{fig:ios:lp}
        \end{subfigure}\hfill
        \begin{subfigure}[t]{0.33\linewidth}
            \centering
            \includegraphics[width=\linewidth, keepaspectratio]{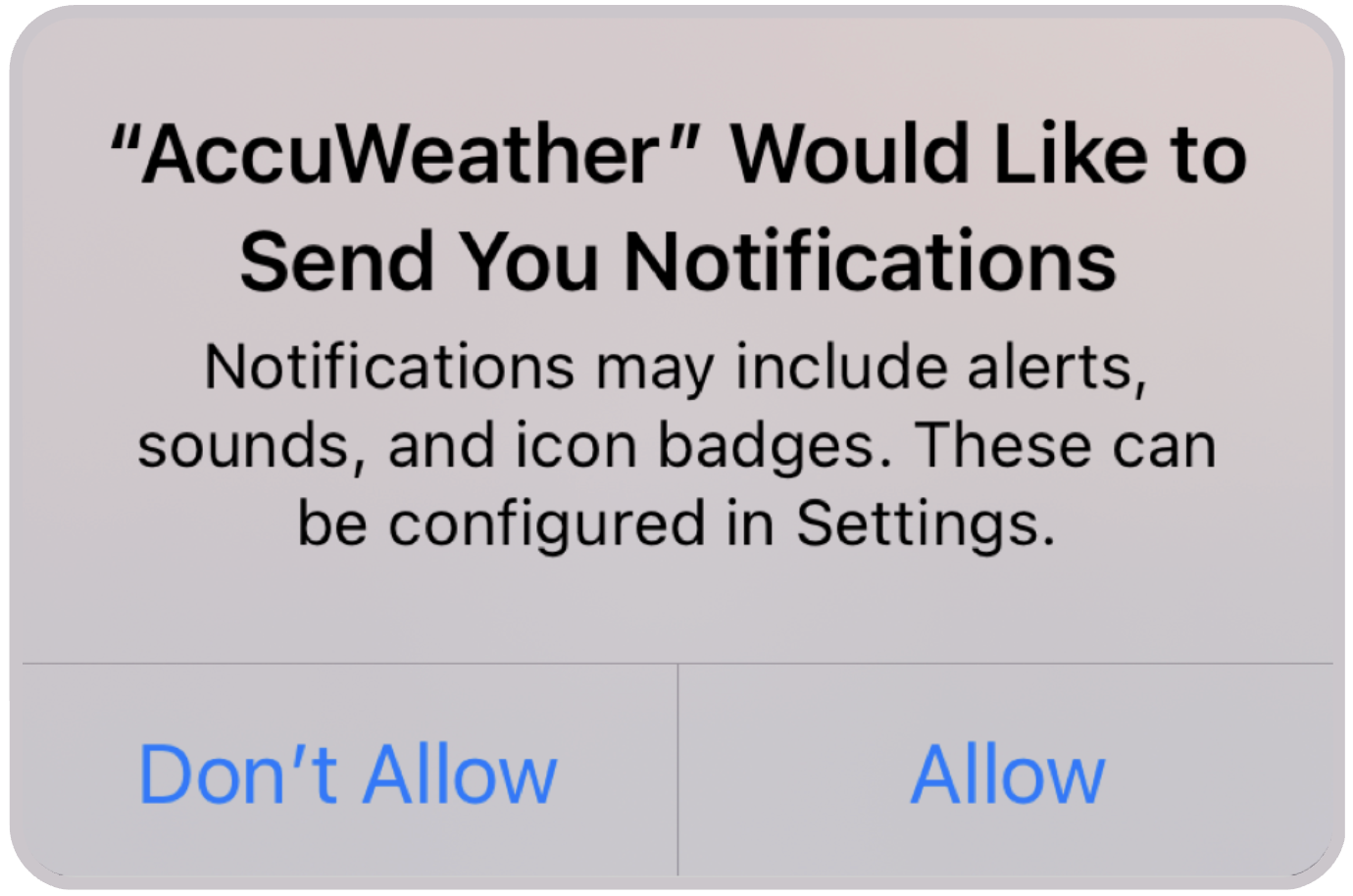}
            \Description{Screenshot of an iOS notification pop-up asking: ""AccuWeather" Would Like to Send You Notifications." Below, the message reads: "Notifications may include alerts, sounds, and icon badges. These can be configured in Settings." Two buttons appear at the bottom: on the left, "Don’t Allow," and on the right, "Allow." The pop-up appears centered on the iPhone screen.}
            \caption{\textbf{NP} ~~ --- ~~ Notification permission}
            \label{fig:ios:np}
        \end{subfigure} 
    \label{fig:ios}
    \end{figure}
}

\newcommand{\spamfigures}{
    \begin{figure}[]
        \centering
        \Description{Three screenshots of spam banners that show up when opening a suspicious Gmail message, described below.}
        \caption{Gmail Spam Notifications}
        \begin{subfigure}[t]{0.48\linewidth}
            \centering
            \Description{Screenshot of gray ``spam'' indicator in Gmail, a gray rectangle between the sender and the email content with text and buttons. The text says, "Why is this message in spam? This message is similar to messages that were identified as spam in the past." The left button is an outlined rectangle stating "Report not spam." There is a small information icon button (circle with the letter "i" in the middle) in the bottom right.}
            \includegraphics[width=0.95\linewidth, height=6cm, keepaspectratio]{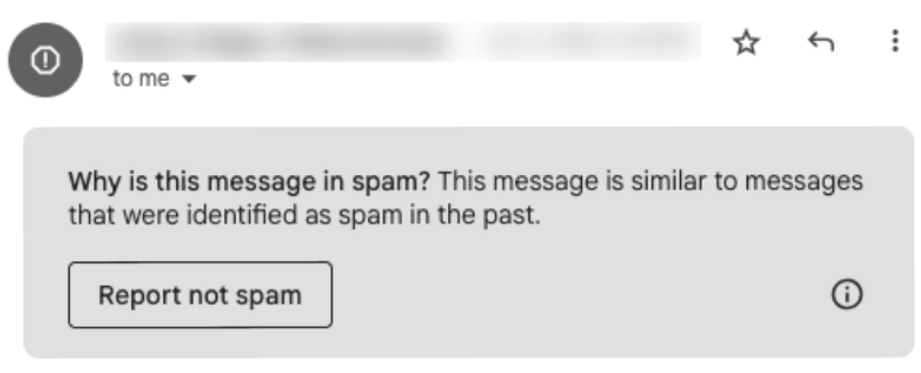}
            \caption{\textbf{SP-G} ~~---~~ Gray "spam" banner}
            \label{fig:spam:spg}
        \end{subfigure}
        \hfill
        \begin{subfigure}[t]{0.48\linewidth}
            \centering
            \Description{Screenshot of red ``malicious content'' indicator in Gmail, a red rectangle between the sender and the email content with text and buttons. The text says, "This message seems dangerous. Similar messages were used to steal people's personal information. Advoid clicking links, downloading attachments, or replying with personal information." The left-most button is an outlined rectangle stating "Looks safe" and there is a small information icon button (circle with the letter "i" in the middle) in the bottom right.}
            \includegraphics[width=0.95\linewidth, height=4cm, keepaspectratio]{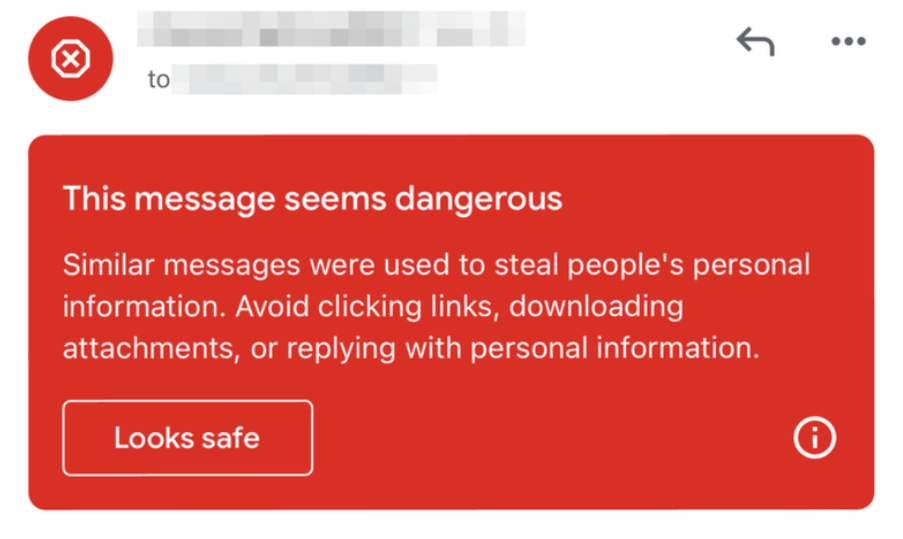}
            \caption{\textbf{SP-R} ~~---~~ Red "malicious content" warning}
            \label{fig:spam:spr}
        \end{subfigure}
        
        \vspace{1em}
        
        \begin{subfigure}[t]{0.9\linewidth}
            \centering
            \Description{Screenshot of yellow ``phishing/external'' indicator in Gmail, a yellow rectangle between the sender and the email content with text and buttons. The text says, "Be careful with this message. [Redacted] is similar to a name in your organization, but the email address does not belong to your domain or [redacted] could not verify that it actually came from [redacted]. Avoid replying to this email unless you contact the sender by other means to ensure that this email address is legitimate." The left-most button is an outlined rectangle stating "Report phishing" and another button with no outline says "Looks safe." There is a small information icon button (circle with the letter "i" in the middle) in the bottom right.}
            \includegraphics[width=0.8\linewidth, height=4cm, keepaspectratio]{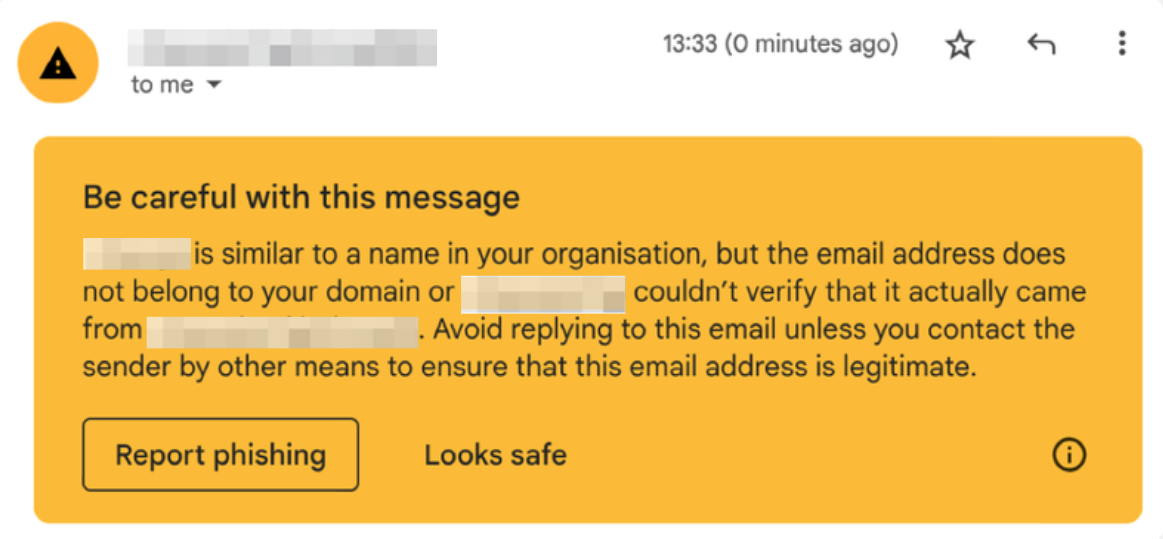}
            \caption{\textbf{SP-Y} ~~---~~ Yellow "phishing/external" email warning}
            \label{fig:spam:spy}
        \end{subfigure}
        
        \label{fig:spam}
    \end{figure}
}

\newcommand{\cookieconsentfigures}{
    \begin{figure}[]
        \centering
        \Description{Two side-by-side screenshots of a cookie consent banner, descriptions below.}
        \caption{Cookie Consent Banner (\textbf{CC})}
        \begin{subfigure}[t]{0.58\linewidth}
            \centering
            \fbox{\includegraphics[height=4.5cm, width=\linewidth, keepaspectratio]{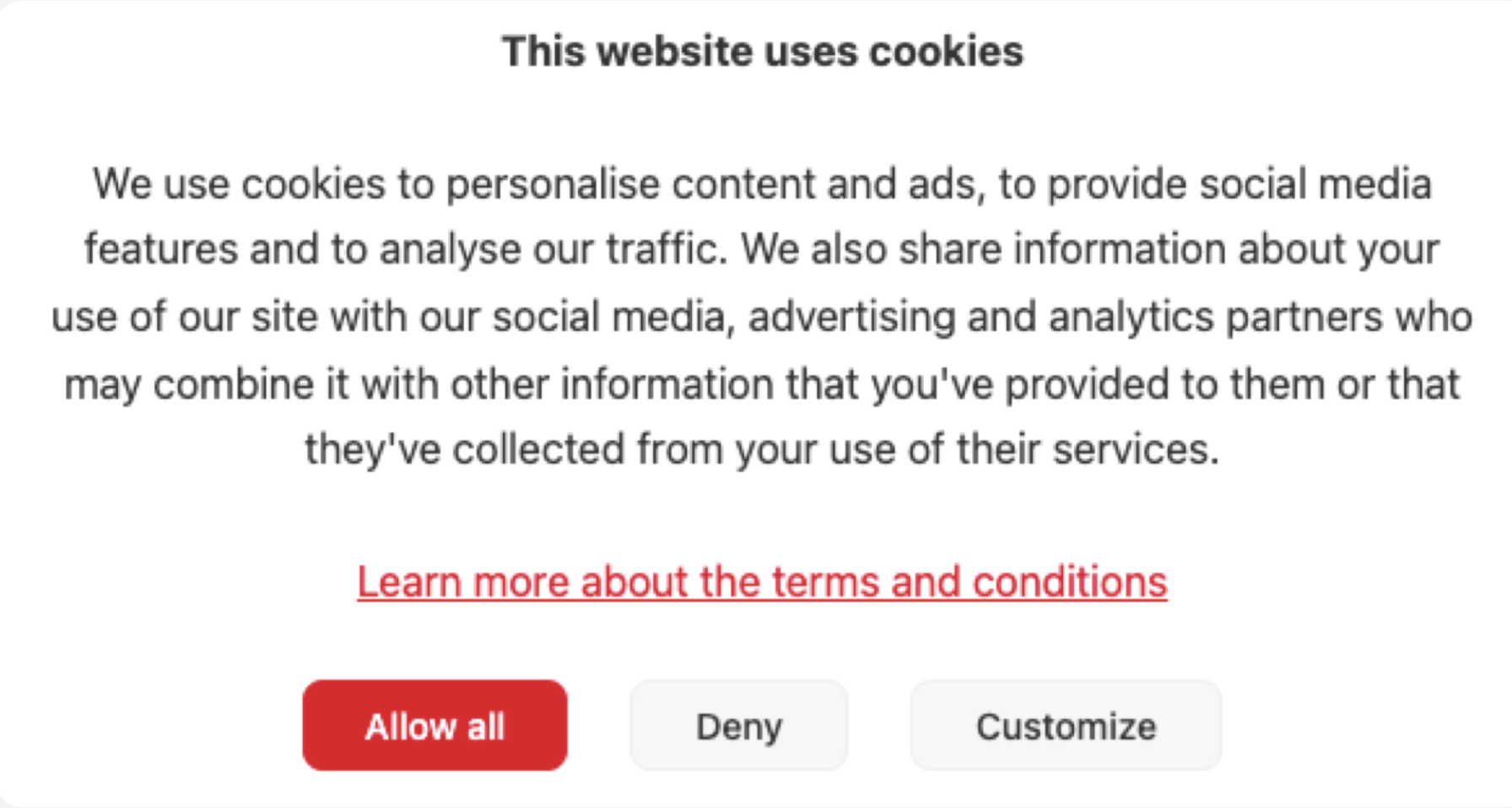}}
            \Description{The left image shows the cookie banner when the user enters the site. From top to bottom, there is text on a white background with a red hyperlink and 3 buttons side-by-side. The text says: "This website uses cookies. We use cookies to personalize content and ads, to provide social media features, and to analyze our traffic. We also share information about your use of our site with our social media, advertising, and analytics partners who may combine it with other information that you've provided to them or that they've collected form your use of their services." The hyperlink states "Learn more about the terms and conditions." The left-most button is a red rectangle and says "Allow All" in white text. The middle button says "Deny" in gray text on a light-gray background. The right-most looks similar with "Customize."}
            \caption{\textbf{CC} when user enters site}
            
            \label{fig:cc:basic}
        \end{subfigure}\hfill
        \begin{subfigure}[t]{0.4\linewidth}
            \centering
            \fbox{\includegraphics[height=4.5cm, width=\linewidth, keepaspectratio]{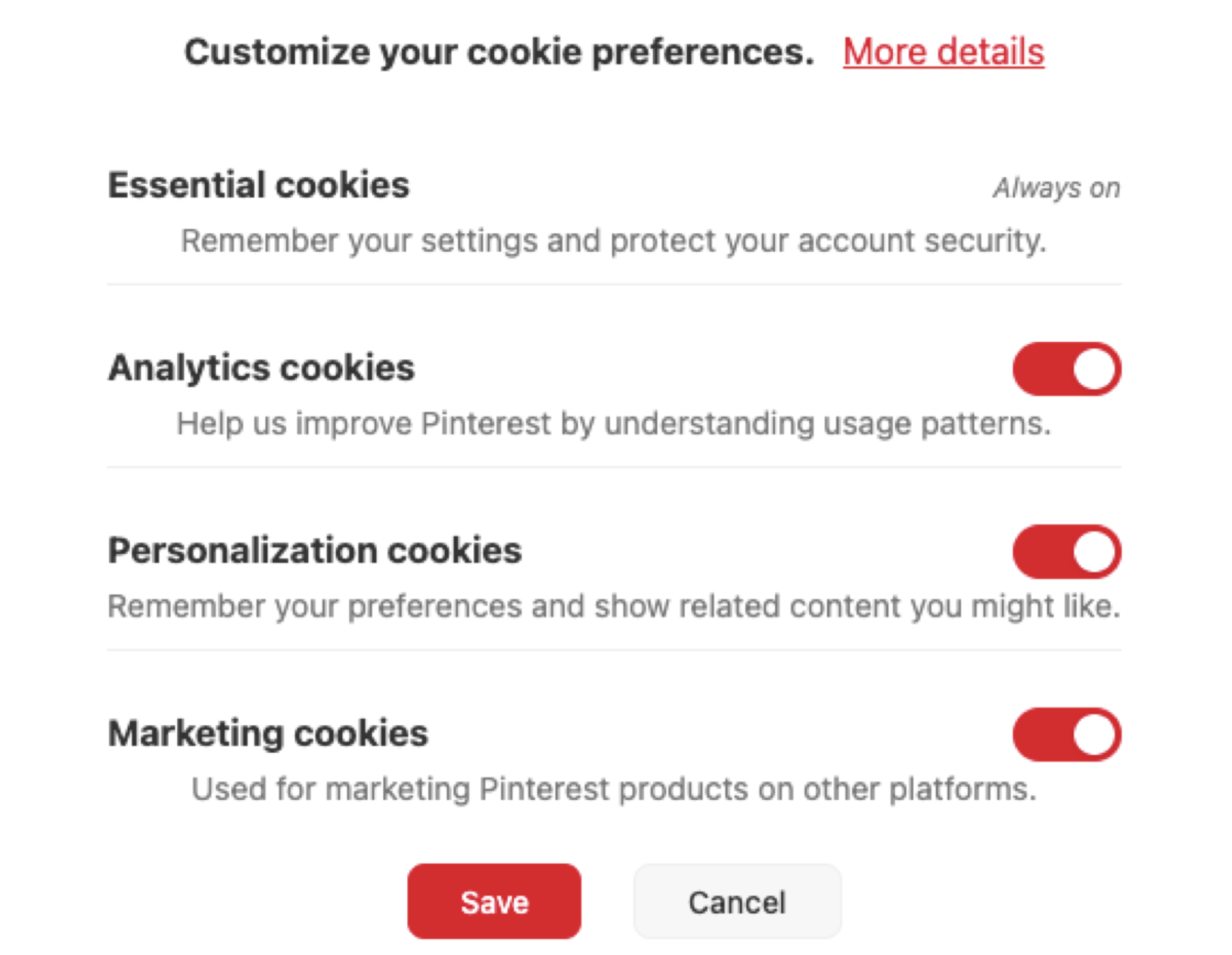}}
            \Description{The image on the right shows a customization menu ("Customize your cookie preferences") where users can enable certain types of notifications in a list with toggle buttons for each (red, horizontal ovals with a circle that moves side-to-side based on whether it is enabled). The four types of cookies listed are: "essential cookies: always on," "analytics cookies," "personalization cookies," and "marketing cookies."}
            
            \caption{\textbf{CC} after selecting ``customize'' option}
            \label{fig:cc:expanded}
        \end{subfigure}
    \label{fig:cc}
    \end{figure}
}

\newcommand{\nsifigure}{
\begin{figure}[]
    \centering
    \Description{This is an email notification. Within the email you see in the top left "Pinterest" and their logo. Reading from top to bottom it says "We noticed a new login." Below that, it has an image of a computer screen. Below that, it has device information in bolded font (MacOS, Chrome, [redacted location]. Below that, it has the date and time of the notification with [redacted time zone]. Below that, it explains that if it was not you, you should secure your account, which is a hyperlink. And finally, at the bottom is a learn more hyperlink about how to keep your account secure.}
    \caption{New Sign-In Email (\textbf{NSI}) ~~ --- ~~ Participants encountered two versions with different login details.}
    \fbox{\includegraphics[height=8cm, width=\linewidth, keepaspectratio]{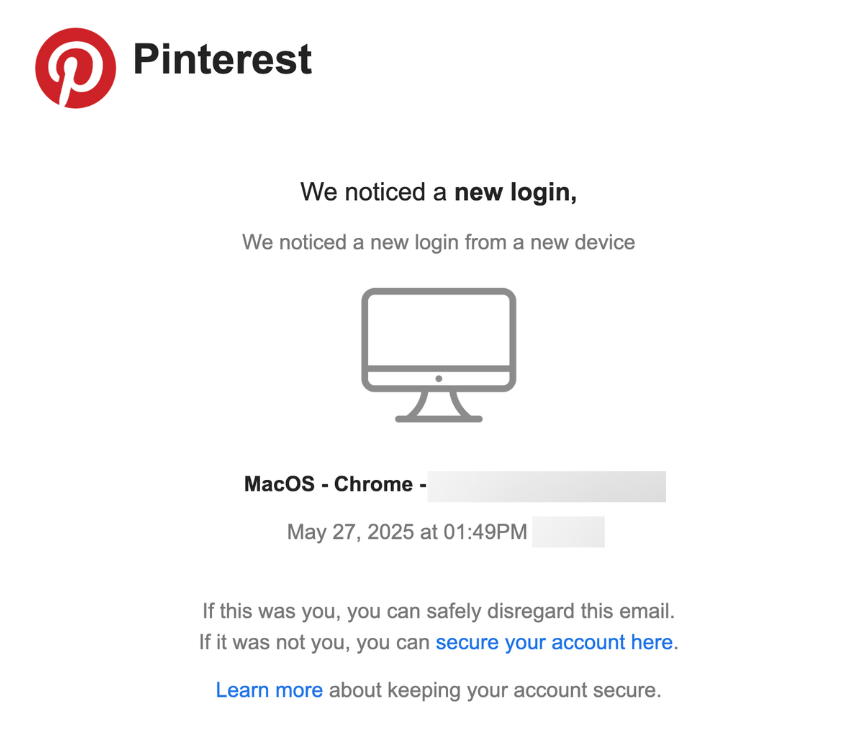}}
    \label{fig:nsi}
    \end{figure}
}      

\newcommand{\srimuifigures}{
\begin{figure}[b]
    \centering
    \Description{Cropped screenshot of an iPhone showing the top centimeter of the home screen where the clock, wifi, and battery symbols reside. Screen recording is enabled, meaning that the clock on the left-hand side is overlaid on top of a rounded, red oval. There is a black annotation arrow pointing to this notification with the label "SRI." Above the Wi-Fi and cellular connection symbol (which says "SOS" as the device is disconnected), there is a small yellow-orange dot. This is the microphone use indicator which turns on if voice recording is ongoing. There is an annotation arrow pointing to this notification with the label "MUI."}
    \caption{Recording Indicators ~~ --- ~~  screen recording indicator (\textbf{SRI}) and microphone use indicator (\textbf{MUI}) on iOS.}
    \label{fig:srimui}
    \includegraphics[width=0.7\linewidth, keepaspectratio]{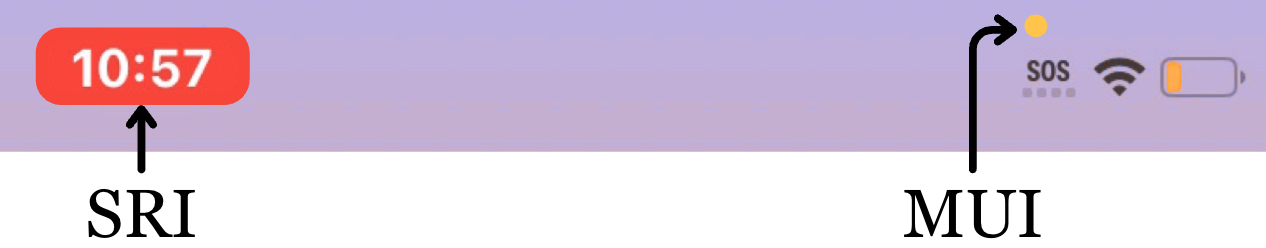}
    \end{figure}
}      

\twoFAfigures
\chromefigures
\cookieconsentfigures
\iospermissionfigures

\srimuifigures
\spamfigures

\par\vspace{15pt}

\nsifigure

\clearpage

\section{Notification-Based Participant Interactions} 
\label{app_interact}

\renewcommand{\arraystretch}{1.1}
\footnotesize
\begin{longtable}{p{2.0cm} p{2.3cm} p{0.5cm} p{10.0cm}}
\caption{Participant actions in response to each security and privacy notification, drawn from the full range of interactions available on screen. Participant reasoning combines behavioral, eye-tracking, and transcription data.}
\Description{A four-column table cataloging participant actions for each of the twelve S\&P notifications used in the study (2FA, MW, HT, CC, TP, NP, LP, SP, DD, NSI, SRI, MUI). For each notification, rows list the possible response categories (e.g., clicking through, avoidance, exploratory clicking, correct or incorrect identification), which participants took each action, and their behavioral and verbal reasoning drawn from video, eye-tracking, and transcript data.}
\label{tab:interact-elements} \\
\toprule
\textbf{Notification} & \textbf{Possible Actions} & \textbf{PID} & \textbf{Reasoning (behavioral + verbal)} \\
\midrule
\endfirsthead

\multicolumn{4}{c}{\tablename\ \thetable{} -- continued from previous page} \\
\toprule
\textbf{Notification} & \textbf{Possible Actions} & \textbf{PID} & \textbf{Reasoning (behavioral + verbal)} \\
\midrule
\endhead

\multirow[t]{26}{2.0cm}{Actionable: 2FA}
 & 2FA-C: ``resend link''
   & None & No participant clicked a dedicated resend button. \\
\cmidrule{2-4}
 & \multirow[t]{1}{2.3cm}{2FA-C: ``change email''}
   & P4 & Clicked the email field/dropdown, which reset her to re-entering her email and password. Leaned back and pet her hair when the 2FA screen reappeared. ``I don't know, why is this saying that again?'' \\
\cmidrule{2-4}
 & \multirow[t]{4}{2.3cm}{2FA-C: Avoidance}
   & P1 & Read the on-screen instructions and buttons but did not connect them to the phone in front of him; leaned back, disengaged. 11-second pause when asked what he'd do. ``Nothing.''  \\
 & & P2 & Gaze bounced across the screen without settling long enough to read anything. Asked for reassurance, in a nervous tone: ``Can I do this one?'' \\
 & & P3 & Looked around, confused expression. Reached for ``try another way,'' then pulled back and grabbed his head (visible distress) before acting when told directly what to do: ``Oh oh oh, open open.'' \\
 & & P6 & Surprised when it appeared; gaze cycled between animation, title, and email icon without landing on an action. Read the title (``two-step verification'') but not instructions. \\
\cmidrule{2-4}
 & \multirow[t]{7}{2.3cm}{2FA-B: Click}
   & P1 & Did not act until redirected (``For the study, this is your phone''), then clicked quickly. \\
 & & P2 & Needed prompting twice before clicking; laughed and admitted confusion about which screen to attend to: ``I didn't know if it was this one or that one.'' \\
 & & P3 & Stared at the phone until told directly to click, then clicked immediately and glanced back at the laptop to check for a change. \\
 & & P4 & Guided to the notification by the researcher. \\
 & & P5 & Moved to the phone quickly. \\
 & & P6 & Quickly leaned in close and tapped immediately, then checked back at the computer screen rather than waiting for 2FA-I. \\
 & & P7 & Needed prompting after the phone's lockscreen went dark on its own; did not wake the phone back up unprompted. \\
\cmidrule{2-4}
 & \multirow[t]{1}{2.3cm}{2FA-B: Avoidance}
   & None & Every participant eventually clicked the banner, though several needed prompting first (see Click, above). \\
\cmidrule{2-4}
 & \multirow[t]{2}{2.3cm}{2FA-I: Exploratory}
   & P6 & Tapped the location text (not a button) before finding and clicking the real yes/no buttons. \\
 & & P7 & Reported tapping the device-info text before self-correcting to the actual ``yes, it's me'' button. \\
\cmidrule{2-4}
 & 2FA-I: ``No, don't allow''
   & None & No participant selected this option. \\
\cmidrule{2-4}
 & \multirow[t]{7}{2.3cm}{2FA-I: ``yes, its me''}
   & P1 & Confirmed quickly (``Yeah'') without appearing to review the identity details shown. \\
 & & P2 & Read for 8 seconds in silence before confirming; never visibly connected the phone-side confirmation to the parallel change on the laptop screen. \\
 & & P3 & Briefly checked buttons and confirmed, then proactively showed the researcher the laptop had updated. \\
 & & P4 & Noticed the resulting screen change immediately. Asked what made her click yes: ``[the `yes, its me' icon] was the green.'' \\
 & & P5 & Looked at the phone for 9 seconds before clicking yes; did not go back to check the laptop screen afterward. \\
 & & P6 & Read the location aloud correctly and confirmed it matched, but tried to tap the location text itself before being redirected to the actual button; eventually clicked ``yes, it's me. ''\\
 & & P7 & Went directly to the app once the phone was unlocked and clicked confidently. \\
\cmidrule{2-4}
 & \multirow[t]{1}{2.3cm}{2FA-I: Back button}
   & None & No participant used the 2FA-I back button to exit this screen. \\
\cmidrule{2-4}
 & 2FA-I: Avoidance & None & All 7 participants confirmed the prompt, several only after hesitation or flow confusion captured above. \\
\midrule

\multirow[t]{11}{2.0cm}{Actionable: Malicious Site Warning (MW)}
 & \multirow[t]{6}{2.3cm}{``Back to safety''}
   & P2 & Face visibly dropped on seeing the warning; glanced at browser tabs before settling in to read. Eye path then moved back and forth between Details and Back to Safety before clicking. ``Now, I'm not sure what to do\ldots Okay, here, uh, back to safety, okay.'' \\
 & & P3 & Surprised (``Oh! H-hello there''); hovered near Details, then pulled back, saying he worried it led to ``the scary thing.'' Second encounter (same notification, later in session) was faster and more direct, straight to Back to Safety despite reacting with surprise again. ``If it was green, I would accept it.'' \\
 & & P4 & Reacted with surprise, pointed at the screen, and named the danger from the title and red color alone before reading anything else (``This says `dangerous site,' the red things.''). Picked Back to Safety once SP4 read out the two button options.  \\
 & & P5 & No visible alarm; clicked quickly and directly, faster than most others. Said, ``No\ldots I don't like that,'' though this seemingly stemmed from the interruption to her preferred activity. \\
 & & P6 & Audible surprise (``Uh!''). Exploratory rather than direct: clicked a large info panel first (nothing happened), then Details, then Back to Safety, without pausing to look closely at any of them. Called it ``bad'' on sight, citing the stop-sign icon as his evidence. \\
 & & P7 & Brief deliberation: looked at the title, then hovered back and forth between Details and Back to Safety before clicking the latter. Emotional response read as suspicion. Read the full warning text fluently once and reported MW ``means going back to safety.'' \\
\cmidrule{2-4}
 & ``Details'' & P6 & Clicked here second in his three-part sequence, right after the larger info panel didn't respond. Moved quickly, without pausing to read the details text. \\
\cmidrule{2-4}
 & ``Learn more about this warning'' & None & No participant used this option.  \\
\cmidrule{2-4}
 & ``proceed to unsafe site'' & None & No participant used this option. P3 named the one condition under which he would: ``If it was green, I would accept it.'' \\
\cmidrule{2-4}
 & \multirow[t]{2}{2.3cm}{Avoidance}
   & P1 & Switched tabs quickly, no visible urgency or alarm. ``I don't really care.'' \\
 & & P2 & Before choosing to read (see ``Back to safety'' above), her expression dropped, and she began looking toward the browser tabs. Possibly continued for the researchers. \\
\midrule

\multirow[t]{10}{2.0cm}{Actionable: HTTPS Warning (HT)}
 & \multirow[t]{5}{2.3cm}{``Back to safety''}
   & P2 & Reacted with mild surprise (raised eyebrows, leaned back), then scanned quickly and clicked. \\
 & & P3 & Eye first drawn to ``Advanced'' due to its left-side position, then moved to Back to Safety. Voice showed concern. Read the warning icon as meaning ``do not\ldots look at it.'' \\
 & & P4 & Hand went to her head immediately on seeing it; P4 and SP4 read the warning text aloud. Chose Back to Safety once her mother read out both button names. \\
 & & P5 & Clicked immediately, though visibly distracted (still engaged with the prior conversation). Called it ``bad'' and said, twice, that it ``makes me throw up.'' \\
 & & P7 & Brief, real reading attempt (read the title, started the body text) before going straight to Back to Safety; looked to researcher and smiled afterward. \\
\cmidrule{2-4}
 & \multirow[t]{1}{2.3cm}{``Advanced''}
   & None & No participant used this option. \\
\cmidrule{2-4}
 & ``Learn more about this warning''
   & None & No participant used this option. \\
\cmidrule{2-4}
 & \multirow[t]{1}{2.3cm}{``proceed to [website] (unsafe)''}
   & None & No participant proceeded. P5 said she would click the continuation hyperlink, then, asked what it does, responded ``I do not [know].'' Her live behavior went to Back to Safety regardless. \\
\cmidrule{2-4}
 & \multirow[t]{2}{2.3cm}{Avoidance}
   & P1 & Responded with low engagement and closed the tab. Calling the site itself ``private.'' Followed by ``could be... dangerous... but I just don't care.'' and that he ``always clicks out.''\\
 & & P6 & Changed tabs back to the Pinterest tab. Called it ``bad''; needed a personal analogy from his mother (someone taking his iPad) before agreeing it mattered. \\
\midrule

\multirow[t]{13}{2.0cm}{Actionable: Cookie Consent (CC)}
 & \multirow[t]{5}{2.3cm}{``Allow all''}
   & P1 & Clicked quickly, without opening Customize or reading the banner; appeared more focused on the Pinterest images behind it. ``I don't read 'em, but I do understand it.'' Believed Allow was letting him use the site, not sharing his data. \\
 & & P2 & Read closely and slowly: glanced between Allow All and the login button twice, then read the body text, over a 35-second pause before answering. Visibly anxious: ``Because you really need to read this\ldots and making sure that it's good.'' Chose Allow because the banner said cookies were needed ``to use our site''; said plainly she didn't know what Deny does. \\
 & & P3 & Scanned across Deny and Allow before asking outright, ``Which one should I pick?'' Landed on Allow via ``a gut feeling,'' denying that the red color swayed him when asked directly. Asked what cookies are, replied ``Eating,'' which R1 confirmed (``Oh, for eating. Yes.''). \\
 & & P4 & First tried to skip the banner entirely by clicking the site's own login button; when it failed, read this as the button being broken rather than the banner needing an answer (see Avoidance, below). Needed nearly every word decoded by her mother before choosing Allow without weighing the alternatives. \\
 & & P5 & Fast, direct click with no visible scanning or hesitation. Asked why, first said ``because it's a picture,'' then agreed when reframed as ``because it's red.'' Read ``cookie'' as the food throughout, describing her own cooking instead of engaging with the prompt. \\
\cmidrule{2-4}
 & \multirow[t]{2}{2.3cm}{``Deny''}
   & P6 & Looked at both Allow All and Deny (never at Customize) before choosing Deny, reasoning ``red = bad'' about the Allow All button's color. Minutes later in the interview, he misremembered and pointed to Allow All as what he'd clicked, until corrected. Read ``cookies'' as ``chocolate chip.'' \\
 & & P7 & Looked surprised (``what's this?''), read only two of the three button labels, and never read the surrounding text (confirmed by eye tracking). Explained Allow accurately in interview but described Deny incorrectly, as losing ads/content entirely. \\
\cmidrule{2-4}
  & \multirow[t]{1}{2.3cm}{``Learn more about terms and conditions''}
   & None & No participant used this option. P6 was asked what the link would do and said ``nothing,'' consistent with a broader pattern of assuming underlined/linked text isn't actually interactive. \\
\cmidrule{2-4}
 & \multirow[t]{1}{2.3cm}{``Customize''}
   & None & No participant clicked into Customize themselves live; all 7 only saw that view once the researcher opened it for them afterward. \\
\cmidrule{2-4}
 & \multirow[t]{1}{2.3cm}{Avoidance}
   & P2 & Asked if she needed to answer the CC and continually looked back at login button, saying ``Okay, um, do I do with this?'' once realizing it was blocking.  \\
 & & P4 & Tried to click the site's login button. Only engaged with the CC (and chose Allow all; see above) once that bypass attempt failed. \\
\midrule

\multirow[t]{8}{2.0cm}{Actionable: Cross-App Tracking (TP)}
 & \multirow[t]{2}{2.3cm}{``Ask App Not to Track''}
   & P2 & Long, hesitant pause (leaned back, visibly confused) before answering. Reasoned via an unrelated real-world habit (avoiding her family's shared Spotify account during her dad's basketball games): ``has to be really careful on what you press.'' \\
 & & P3 & Read the on-screen AccuWeather text aloud first, then stated his decision separately: ``you don't need to track.'' Reasoned that not wanting ads meant declining tracking. \\
\cmidrule{2-4}
 & \multirow[t]{5}{2.3cm}{``Allow''}
   & P1 & Clicked Allow immediately, without reading, describing it as his habit for apps he considers necessary. Associated ``tracking'' with Find My iPhone rather than advertising. \\
 & & P4 & Paused noticeably before choosing Allow, on an app-updated version of this prompt. Explaining ``track'' as location-tracking incorrectly. \\
 & & P5 & Explored through ``customize'' toggles (updated TP) and several more screens quickly to reach the weather content, treating the screens as obstacles rather than decisions. Guessed the notification was ``for your phone number.'' \\
 & & P6 & Chose ``customize'' on the updated in-app screen, then clicked through several more screens (unit settings, map style, next) toward an account-creation step he was told he could skip. A plain-text ``Maybe Later'' link did not register to him as clickable. \\
 & & P7 & Clicked Allow (update TP), contradicting her own interview-stated preference to decline tracking (see Avoidance, below). \\
\cmidrule{2-4}
 & \multirow[t]{1}{2.3cm}{Avoidance}
   & P7 & Before answering, tried clicking outside the permission dialog to make it disappear without choosing either option, only picking Accept once she realized a choice was required. Said she doesn't know what is being tracked and where (``I don't even know.'')  \\
\midrule

\multirow[t]{8}{2.0cm}{Actionable: Notification Permission (NP)}
 & \multirow[t]{1}{2.3cm}{``Don't allow''}
   & P2 & Clicked during the task; in interview, was unsure she remembered the choice correctly (``I did say don't allow\ldots I wasn't as sure, I was going back and forth''), suggesting the live decision may not have been highly deliberate. On notifications generally: ``I like [notifications], but sometimes it can be annoying at times\ldots if you have too many of them.'' \\
\cmidrule{2-4}
 & \multirow[t]{6}{2.3cm}{``Allow''}
   & P1 & Clicked Allow immediately without reading. Framed it as a gate to ``access'' the app. \\
 & & P3 & Read the app's own tagline aloud first (``Hmm stay a-ahead of-f-f storm.''), then asked researchers to choose; only confirmed wanting weather notifications once NP's purpose was explained to him. \\
 & & P4 & Fast click on the bottom button. Misread ``notifications'' as ``locations'' in interview, and her justification for Allow stayed location-based even after SP4 corrected the misreading. \\
 & & P5 & Immediate, confident click; said ``Yeah'' when asked if she always says allow. Says she doesn't like getting notifications, but allows them every time. \\
 & & P6 & Said he knew what a notification was, understanding after SP6 gave a concrete example (``when it goes ding!''). Confirmed liking notifications. \\
 & & P7 & Explained that she keeps notifications on for weather for regular updates; reported disabling them for some other apps situationally. \\
\cmidrule{2-4}
 & \multirow[t]{1}{2.3cm}{Avoidance}
   & None & No participant avoided answering this prompt. \\
\midrule

\multirow[t]{9}{2.0cm}{Actionable: Location Permission (LP)}
 & \multirow[t]{1}{2.3cm}{Choice not logged}
   & P1 & Paused noticeably longer before choosing than he had on TP/NP, which the researcher flagged as possibly feeling more consequential to him. Which specific button he chose was not recorded in the underlying data. \\
\cmidrule{2-4}
 & \multirow[t]{3}{2.3cm}{``Allow once''}
   & P5 & Described location-tracking as ``they track your home.'' Explained the three buttons spatially rather than by content: ``allow once'' takes you ``to the app,'' ``while using'' ``goes up,'' ``don't allow'' ``goes down.'' \\
 & & P6 & Broke his more common ``click the bottom option'' default. Could correctly read all three choices, but misremembered which he'd actually picked (first pointed to ``Don't Allow'' before correction), and could not explain the difference between Allow Once and Don't Allow, guessing (``maybe this one?''). \\
 & & P7 & Self-corrected mid-sentence from ``your location'' to ``my location.'' Could recall she'd chosen Allow Once but not explain why over ``Allow While Using.'' \\
\cmidrule{2-4}
 & ``Allow while using app''
   & None & No participant chose this option. \\
\cmidrule{2-4}
 & \multirow[t]{3}{2.3cm}{``don't allow''}
   & P2 & Framed the decision around her own ``physical privacy''; the only app she knowingly allows location for is Find My. Also declined ``Allow Once,'' reasoning it would mean ``you get all notifications.'' \\
 & & P3 & Reacted with mild surprise and asked directly what the prompt was asking (``What this say now?''), before choosing to deny. Said he'd consider ``Allow Once'' an acceptable middle ground. \\
 & & P4 & Asked ``Why does this say don't allow?'' before still choosing it. Read ``tracking'' as being about the weather, not her own location, even when redirected twice. \\
\cmidrule{2-4}
 & \multirow[t]{1}{2.3cm}{Avoidance}
   & None & No participant avoided answering LP. \\
\midrule

\multirow[t]{9}{2.0cm}{Informational: Spam Banners (SP)}
 & \multirow[t]{1}{2.3cm}{``Report not spam''}
   & None & No participant selected this option.  \\
\cmidrule{2-4}
 & \multirow[t]{1}{2.3cm}{More information (i)}
   & None & No participant selected this option.\\
\cmidrule{2-4}
 & \multirow[t]{3}{2.3cm}{Did not notice in email}
   & P5 & Could not identify any of the three banners when shown them in interview. Gray: offered only an associative guess about streaming content (``videos''). Yellow: first said she didn't know it either, then guessed ``lock it''/``log in'' once pressed. Red: named it most dangerous, but gave no supporting reasoning beyond the color (``that's all''). \\
 & & P6 & Went straight into the flagged email and did not look at SP. Clicked the phishing link. Guessed ``spam'' meant ``maybe super hero?''; rated yellow banner ``safe''/``good''; confidently identified red as ``bad,'' citing the X-mark icon. \\
 & & P7 & Opened and read the flagged email but, per eye-tracking, never looked at the gray banner, though she reported in interview that she had seen it. Interview: read gray fluently and, unlike most participants, correctly explained what ``report not spam'' does even though she's never used it; had no recognition of yellow or the term ``phishing,'' but still produced an cautious response (``be careful'') and ranked it more dangerous than gray; gave a three-tier ranking (red $>$ yellow $>$ gray), reasoning through direct word-matching on red's own text (``it seems bad\ldots `cause it's bad''). \\
\cmidrule{2-4}
 & \multirow[t]{4}{2.3cm}{Avoidance of Spam}
   & P1 & Did not open, delete, or otherwise act on the flagged email until the researcher asked him to open it for discussion. Read gray as purely informational (``I'm just telling you, if it's spam or not'') and said the color ``doesn't matter'' to his own risk assessment; for yellow, judged by sender/content relevance rather than the banner; for red, explicitly rejected color as a signal for himself (``Nothing to me''), while noting it ``mean[s] some'' to SP1. \\
 & & P2 & Chose to delete spam (``I would get rid of it\ldots just in case''). Interview: Stated gray ``makes it boring,'' wanting its warning text made bolder; had little to say about yellow; for red, proposed pairing ``dangerous'' (describing the message) with ``cautious'' (describing the reader's required action). \\
 & & P3 & Deleted the spam email using their ``just get rid of it'' strategy for anything in Spam, confirmed as a standing household rule (ask a parent if unsure). Interview: misread ``report not spam'' as a statement rather than a button; yellow meant nothing to him until SP3 offered a stoplight analogy, which he then applied confidently; red produced immediate, emphatic recognition of severity (``this is very important''), naming color, wording, and the stop icon together as reinforcing cues (``it's all three''). \\
 & & P4 & Called the spam email ``a garbage one'' and deleted it after SP4 helped her sound out the sender name and catch the misspelling (``Phinterest,'' not ``Pinterest''). Interview: consistently read gray as ``there's been hacker spam emails'' and requested a different color to draw attention; recognized yellow immediately (``someone could hack into your email account''); quickly named red ``definitely'' the most dangerous. \\
\midrule

\multirow[t]{12}{2.0cm}{Informational: Dangerous Download (DD)}
 & \multirow[t]{1}{2.3cm}{Click $\rightarrow$ ``Delete from history''}
   & None & No participant clicked delete. P4 and P5 were both offered this in the expanded view and explicitly declined it (P4: would ``ignore it, and just don't look at'' it; P5: chose ``keep it'' over deleting). \\
\cmidrule{2-4}
 & \multirow[t]{1}{2.3cm}{Click $\rightarrow$ ``Learn why Chrome blocks some downloads''}
   & P1 & Leaned in when DD appeared. Clicked into ``Learn more'' rather than dismissing it; read the title, said ``Yep,'' and closed it without further action. Called it a ``blocked site,'' distinguishing DD as something ``okay to\ldots click.'' \\
\cmidrule{2-4}
 & \multirow[t]{1}{2.3cm}{Did not see}
   & P6 & On a re-triggered showing (his second live encounter), did not notice the notification at all, even leaning and moving his head to find it before agreeing it was ``very small.'' His first encounter, he did see it and let it disappear (see Avoidance, below). \\
\cmidrule{2-4}
 & \multirow[t]{3}{2.3cm}{Re-clicked link that triggered notif.}
   & P2 & Clicked trigger button again after the first warning auto-dismissed. Sighed the second time and again let it disappear. \\
 & & P3 & Encountered it twice live, letting it auto-dismiss both times, then intentionally re-triggered it once more himself to show the researcher what he'd seen. \\
 & & P4 & Clicked trigger button six times before the notification even appeared, because its slight loading delay made her think the button ``wasn't letting'' her proceed. \\
\cmidrule{2-4}
 & \multirow[t]{6}{2.3cm}{Avoidance}
   & P2 & Reacted with an under-the-breath ``interesting,'' then let it auto-dismiss both times without interacting. Explained her non-interaction as risk-averse: believes clicking the notification itself is the dangerous action (``it can open up different\ldots viruses''). \\
 & & P3 & Reacted with visible surprise each time; focused on the word ``blocked'' over ``dangerous.'' Said in interview he'd choose delete if asked, though his actual live behavior both times was to do nothing. \\
 & & P4 & Leaned back and said ``it's dangerous'' with visible concern, then let it disappear without further action. Later generalized the notification's meaning to ``all the pop-ups are now blocked.'' \\
 & & P5 & Reacted (raised eyebrows, small head shake) before continuing. Read the octagon-with-X icon as meaning ``don't do it.'' \\
 & & P6 & On his first encounter, glanced briefly and moved on without interacting (his second, non-noticing encounter is under Did not see, above). Asked what you can do with it, said simply ``nothing.'' Reports not knowing what ``download'' or ``delete from history'' mean. \\
 & & P7 & Watched it with a furrowed brow until it disappeared on its own, then moved to a different image entirely. Explained it as ``trying to get your information,'' i.e., as itself the threat. \\
\midrule

\multirow[t]{8}{2.0cm}{Intrusion: New Sign-In Alert (NSI)}
 & \multirow[t]{3}{2.3cm}{Correct identification}
   & P3 & Used the mismatched city/state in the subject line as their primary signal (``this one I am in\ldots this one I am not in''), and noticed the fraudulent email's unfamiliar OS (``Mac'') without weighing that mismatch as strongly as the matching one (``Windows''). \\
 & & P4 & Got it wrong on her first pass (reviewing only surface logos/icons), then self-corrected mid-conversation once SP4 had her compare the two emails side by side and she started reading the subject line for city names. Did not independently catch the OS or browser mismatches even with repeated prompting; eventually caught the date discrepancy, but only after substantial repeated prompting. \\
 & & P7 & Correctly and quickly identified the fraudulent email by its mismatched city (``the one from [correct city]''), though she did not check OS, browser, or full date/time details even when prompted to look for more. \\
\cmidrule{2-4}
 & \multirow[t]{2}{2.3cm}{Incorrect identification}
   & P1 & Confidently said both emails were his own, reasoning that he'd lived in both listed states so both addresses felt personally valid. Never checked the date/time or considered what the alert was attached to. \\
 & & P5 & Initially starred the email with the recognizable city name in its subject, a correct-leaning first move. When asked to judge each email individually, said yes to both, with unrecognized reasoning (``you had to take a picture of your face\ldots smile for the camera''). \\
\cmidrule{2-4}
 & \multirow[t]{2}{2.3cm}{Indecision}
   & P2 & Read the correct email's city/state right away, then reversed to calling it ``spam,'' then reversed again, repeating this back-and-forth across both emails; at one point treated the embedded ``secure your account'' link as automatically suspicious regardless of which email it was in. Eventually landed back on the correct, location-based answer with visible relief: ``Yeah\ldots Thanks.'' \\
 & & P6 & Opened the fraudulent email first, clicked several non-interactive elements (icon, date/time) that didn't respond, and flipped his answer more than once based on that lack of response rather than on content. Only slowed down and used the correct city/state signal once SP6 prompted (``Do you live in [City]?''). \\
\cmidrule{2-4}
 & \multirow[t]{1}{2.3cm}{``Secure my account''}
   & P7 & Paused for eleven seconds, then said ``secure my account'' and actually hovered over and clicked the link.  \\
\midrule

\multirow[t]{7}{2.0cm}{Intrusion: Screen Recording Indicator (SRI)}
 & \multirow[t]{7}{2.3cm}{Did not notice}
   & P1 & Did not report noticing it during the study; asked about it directly, correctly identified it: ``It's just recording.'' \\
 & & P2 & Guessed the SRI meant the time itself was frozen or ``blocked,'' based on location and color. Proposed a separate dedicated icon non-overlapping with the time. \\
 & & P3 & Did not report noticing it; told directly what it meant, and when asked to recall it later, said ``recording.'' Preferred a design that kept the time visible alongside the indicator over an abstract concentric-circle icon. \\
 & & P4 & Guessed it meant location, then the weather, each guess tracking whatever topic had just been discussed rather than the SRI itself. \\
 & & P5 & Could not explain the indicator verbally beyond tautological answers (``because it's red''), but went into the phone's settings and turned off screen recording, showing she knew what it was. \\
 & & P6 & Guessed it related to the time display it was drawn around (``maybe it's three?''), then, shown an alternate design, described it only by shape (``that's an `O'\ldots maybe a donut''). \\
 & & P7 & No guess offered on first exposure (``no''). Asked again later in the same interview, guessed it was camera-related. \\
\midrule

\multirow[t]{7}{2.0cm}{Intrusion: Microphone Use Indicator (MUI)}
 & \multirow[t]{7}{2.3cm}{Did not notice}
   & P1 & Associated the dot with a workplace availability/status indicator (like Microsoft Teams). \\
 & & P2 & Offered no independent guess before being informed. \\
 & & P3 & Guessed ``wifi'' first, reading the unfamiliar orange dot through its visual proximity to the actual wifi icon. \\
 & & P4 & Guessed ``the Wi-Fi's on, or the internet's on.'' Once corrected, they wanted the indicator removed, but reversed to ``leave it alone'' once its protective purpose was explained. \\
 & & P5 & Session-end fatigue; response uninterpretable (she asked to stop and go home shortly before this notification). Excluded from Appendix~B. \\
 & & P6 & Guessed ``for charger,'' a low-battery association grounded in his own device habits, without cross-checking the phone's actual battery icon. SP6 said she had never known what it meant. \\
 & & P7 & Shook her head, indicating no recognition; guessed ``somebody's spying on you?'' when asked. \\
\bottomrule
\end{longtable}

\section{Notification-Based Participant Design Preferences from Interview Discussion} 
\label{app_design}

\renewcommand{\arraystretch}{1.1}
\begin{longtable}{p{1.7cm} p{2.0cm} p{1.5cm} p{1.3cm} p{8.0cm}}
\caption{Design elements participants found helpful or unhelpful per notification. \textit{Helpful} indicates the element aided the participant's decision and resulted in a safe or appropriate choice; \textit{Unhelpful} indicates it caused misinterpretation or contributed to an unsafe choice. Confusion caused by elements outside the notification and comprehension attributable to prior experience are excluded.}
\Description{A five-column table cataloging which design elements (language, information presented, iconography, color, animation, buttons, and interactive elements) participants found helpful or unhelpful for each of the twelve S\&P notifications used in the study. For each notification and design element, rows indicate whether it was helpful or unhelpful, which participants it applied to, and supporting behavioral and verbal evidence.}
\label{tab:design-elements} \\
\toprule
\textbf{Notification} & \textbf{Design Element} & \textbf{Help Level} & \textbf{PID} & \textbf{Evidence (behavior + quotes)} \\
\midrule
\endfirsthead

\multicolumn{5}{c}{\tablename\ \thetable{} -- continued from previous page} \\
\toprule
\textbf{Notification} & \textbf{Design Element} & \textbf{Help Level} & \textbf{PID} & \textbf{Evidence (behavior + quotes)} \\
\midrule
\endhead

\multirow[t]{7}{1.7cm}{2FA}
 & Language & Unhelpful & P2, P3, P4, P5, P6 & P2 said the instructions needed to be spelled out more. P3 was confused from ``the beginning.'' P4 read ``time sensitive'' only through ``sensitive,'' and related it to the email's sensitivity. P5 read the same phrase only as urgency to log in: ``You have to get in... to email.'' P6 read it but could only guess at its meaning: ``Maybe you ten.'' \\
\cmidrule{3-5}
 &  & Helpful & P3 & Said the ``confirm whether it's you trying to sign in'' phrasing was already clear and needed no more instruction. \\
 \cmidrule{2-5}
 & Information\newline Presented & Helpful & P2 & P2 liked the identity-confirmation details on 2FA-I, specifically the ``just now'' timestamp. \\
\cmidrule{3-5}
 &  & Unhelpful & P2, P4, P5, P7 & P2 objected to the truncated email address on the banner. P4, asked to compare the location text to a map picture showing the same location, said she'd prefer the picture. P5 asked if anything besides the ``yes, it's me'' button was helpful, and said no. P7 said she never looks at the location/device details before choosing, since she just clicks yes reflexively: R1: ``this stuff maybe isn't very helpful to you?'' P7: ``No.'' \\
\cmidrule{2-5}
 & Iconography & Helpful & P3, P4, P5, P6 & All named the green checkmark/red-X pair specifically as what worked for them. P3: easy to use. P4: ``It was the green'' that prompted her to click. P5: ``I like the checkmark.'' P6 read the buttons through their symbols alone (``It's X... it's yes'') and rated the whole screen ``Easy.'' \\
\cmidrule{2-5}
 & Animation & Unhelpful & P2, P3, P4, P6 & P2 found the multi-device animation confusing because she was looking at a computer while it showed a phone, and never connected it to opening the phone: ``It was a little confusing to see the computer and the phone.'' P3 looked back and forth between the animation and the buttons, visibly confused, and could not determine from it that he needed to pick up his phone. P4 found the illustrated phone-opening-Gmail animation confusing rather than clarifying: ``It was confusing,'' and said it ``would have been more real if it was real.'' P6 likewise kept looking back at the animation searching for an answer and needed to be prompted for help rather than resolving his confusion from it alone. \\
\cmidrule{2-5}
 & Buttons & Unhelpful & P2, P6, P7 & P2 worried the ``No, don't allow'' button's position invites accidental clicks. P6 struggled to tell actual buttons apart from static text, repeatedly tapping the location text and timestamp instead of the Yes/No buttons. P7 likewise first named a non-button information field (``Device?'') when asked which button she'd clicked, before correcting herself to ``yes, it's me.'' \\

\midrule
 
\multirow[t]{6}{1.7cm}{MW}
& Language & Unhelpful & P2, P4, P6, P7 & P2 and P4 found the Details view's vocabulary unclear even with help. P6 could read ``Dangerous site'' aloud but did not know what ``dangerous'' itself meant. P7 could fluently read the entire Details paragraph aloud (attackers, stolen passwords, credit card numbers), but when asked what it meant, only repeated ``it means going back to safety.'' Her comprehension appeared to stop at the action rather than the underlying risk. \\
\cmidrule{3-5}
 &  & Helpful & P1, P2, P4, P6, P7 & P1 produced a correct, basic paraphrase unprompted. P2 and P4 both singled out the same phrase as clearest. P6 also read the title correctly on his own: ``Dangerous site.'' P7 responded to the title fastest and most confidently of anyone: ``Dangerous site... I just get out of it really quick.'' \\

 \cmidrule{2-5}
 & Iconography/Color & Helpful & P2, P3, P4, P5, P6 & P2 approved of the red warning page once zoomed in. P3 stated a clear color-conditional rule. P4 named the danger from color and title alone. P5 called it ``bad'' on sight. P6 reacted instantly to the red banner (``Oh, it's bad'') and pointed straight to the stop-sign icon as the reason. \\
 \cmidrule{2-5}
 & Buttons & Unhelpful & P1, P2, P3, P4, P6 & None got functional value from Details. P1 and P4 never clicked it at all. P2 worried it led somewhere risky. P6 first tried clicking a non-button panel before finding Details, then clicked through without appearing to read any of it. \\
 \cmidrule{3-5}
 &  & Helpful & P2 & Named Back to Safety as what she liked most, tying it to asking a parent when unsure: ``What I like about it is... I liked... back to safety... kids will not know, because... they would say, like, um, `dad?''' \\
 \cmidrule{2-5}
 & Interactive \newline Elements & Helpful & P7 & With prompting, correctly identified the ``this site is unsafe'' hyperlink in the Details text as clickable. The only participant to recognize a non-button link by function; also found and used the ``secure my account'' hyperlink on NSI. \\

\midrule
 
\multirow[t]{5}{1.7cm}{HT}
 & Language & Unhelpful & P1, P4, P6 & P1 described the site itself as ``private,'' inverting the warning's intended meaning. P4 misread the word outright: ``Your connection is not provided'' (for ``private''). P6 needed nearly the entire warning read to him word by word. \\
 \cmidrule{3-5}
 &  & Helpful & P3, P7 & P3, in the Advanced view, correctly zeroed in on the one word that mattered: ``it's expired.'' P7 read the full warning fluently and derived the correct cautionary conclusion from it (``you shouldn't click on it''), and could distinguish it from MW when asked to compare the two: ``Both of them are dangerous, but this one is not private.'' \\
 \cmidrule{2-5}
 & Iconography & Helpful & P3, P5, P6 & All three said the triangle-with-! icon stood out to them. P3 said it was more salient here than on MW's page. P5 pointed to it and to the red X in the URL bar. P6 pointed to it unprompted as what showed the page was ``bad.'' \\
\cmidrule{2-5}
 & Buttons & Unhelpful & P1, P3, P6 & None ever clicked through. P1: ``No.'' P3, asked if he'd ever proceed to the site: ``No.'' P6 showed full avoidance too, never touching a button and going straight back to the Pinterest tab. \\
\cmidrule{2-5}
 & Interactive \newline Elements & Unhelpful & P5 & Said she would click the ``continue'' hyperlink despite not knowing what it does: ``Yes'' she would click it; ``I do not'' know what it does. \\

\midrule
 
\multirow[t]{3}{1.7cm}{CC}
 & Language & Unhelpful & P1, P2, P3, P4, P5, P6, P7 & P1 described Cookie Consent as simply ``allowing you to do something.'' P2 was unsettled by ``share information.'' P3 read ``cookies'' literally as food to eat. P4 needed nearly every word sounded out for her. P5 read ``cookies'' through cooking. P6 read ``cookies'' as ``chocolate chip.'' P7 read Deny as meaning she'd get ``any social media and ads and content,'' an inverted, overbroad understanding of what the setting actually restricts. \\
\cmidrule{2-5}
 & Buttons & Unhelpful & P2, P3, P4, P5, P6, P7 & None chose their button through comprehension of what it does. P2 did not know what Deny would do. P3 chose Allow through ``a gut feeling.'' P4 clicked Allow without weighing alternatives. P5 and P6 both chose purely by color. P7 chose Deny for the reason above, but on the Customize screen moments later said to leave every cookie category on, an inconsistency she could not explain. \\
\cmidrule{2-5}
 & Interactive \newline Elements & Unhelpful & P1, P2, P3, P4, P6 & None engaged meaningfully with the Customize toggles. P1 showed no recognition of the view. P2 avoided it live. P3 toggled through without apparent understanding. P4 conflated the toggles with notification settings. P6 did not recognize the toggles as interactive at all and thought clicking the Terms and Conditions link would do ``Nothing.'' \\
\midrule
 
\multirow[t]{1}{1.7cm}{TP}
 & Language & Unhelpful & P1, P2, P3, P4, P5, P6, P7 & P1 ended confused about whether the button labels map onto the tracking concept at all. P2 extended the permission's scope to other companies and shared accounts. P3 equated tracking only with ads. P4 defined ``track'' differently across notifications. P5 guessed it was ``for your phone number.'' P6 needed nearly every word of the banner read to him. P7 read the banner fluently but admitted total non-comprehension of what was being tracked or by whom: ``I don't even know,'' and ``No'' when asked if she knew what ``track your activity'' meant. \\
\midrule
 
\multirow[t]{1}{1.7cm}{NP}
 & Language & Unhelpful & P4, P5, P7 & P4 read ``notifications'' as ``locations'' while reading aloud. P5 could not read the word ``notifications'' at all. P7 needed ``configured'' read for her, misreading it as ``confined.'' \\
\midrule
 
\multirow[t]{3}{1.7cm}{LP}
 & Language & Unhelpful & P3, P4 & P3 did not understand the prompt's purpose from its own text. P4 read ``tracking'' as being about the weather rather than her own location. \\
\cmidrule{3-5}
 &  & Helpful & P2 & Used the permission's own wording to reach a specific, reasoned decision (``just give me a minute by minute weather updates, but wherever you are''), and declined for privacy. \\
\cmidrule{2-5}
 & Buttons & Unhelpful & P1, P2, P3, P5, P6, P7 & None could distinguish the tiers by actual scope. P1: ``it's all the same.'' P2 declined ``Allow Once'' believing it meant ongoing alerts. P3 could not explain why he'd reject ``Allow While Using.'' P5 described all three buttons spatially. P6 misread the labels outright. P7 read the labels correctly this time but still could not explain her own choice: ``I don't even know why I didn't click [Allow While Using]... but--'' \\
\midrule
 
\multirow[t]{9}{1.7cm}{SP}
& Language & Unhelpful & P2, P3, P4, P5, P6, P7 & P2 flagged ``seems dangerous'' as too soft. P3 did not know ``phishing.'' P4 needed ``automatically'' and ``identify'' sounded out. P5 did not know the gray or yellow banners at all. P6 needed nearly every word read for him. P7 also did not know the word ``phishing'' at all: ``No.'' \\
\cmidrule{3-5}
 &  & Helpful & P7 & Correctly explained what the ``report not spam'' button's logic was just from reading it, despite never having used it herself: ``So I, if I don't think it's spam, report it.'' \\
 \cmidrule{2-5}
 & Iconography & Unhelpful & P1, P2, P3, P4, P5, P6 & None understood the (i) icon. P1 said he'd never click it. P2 misread it as itself a warning. P3 didn't know what it meant. P4 guessed ``deleted'' then ``eye.'' P5 guessed streaming videos. P6 guessed a question mark, then a period, and ultimately called it ``a picture'' rather than something clickable. \\
\cmidrule{2-5}
& Color & Unhelpful & P6 & Gave contradictory color judgments across the tiers: correctly called red ``bad,'' but called yellow ``safe'' even compared directly against red, and later called a red stop-sign icon ``good'' several times before self-correcting. \\
\cmidrule{3-5}
 &  & Helpful & P3, P4, P5, P7 & All read the color tiers correctly using color alone. P3: red combines with wording and icon. P4 ranked all three correctly when asked directly (``Definitely\ldots the red one'' as most dangerous). P5 named red as most dangerous with no other reasoning needed. P7 gave the fullest account of any participant, articulating a distinct rule for each tier: gray means ``you can open it and see what it is, and then if you don't like it, delete it''; yellow means ``be careful''; red means ``dangerous.'' \\
\cmidrule{2-5}
 & Buttons & Unhelpful & P5, P6 & P5 misread the ``Looks Safe'' button label (transcribed variously as ``lock it'' or ``log in,'' unclear which; ``block'' for a different tier). P6 was the only participant to actually click the spam email's malicious link, doing so twice out of curiosity about what it would show. \\

\cmidrule{2-5}
 & Interactive \newline Elements & Unhelpful & P2, P3, P4 & None recognized non-button clickable text. P2 did not see the blue ``click here'' text as clickable. P3 read the same style of text as a passive statement. P4 read ``delete all spam messages now'' as a factual statement rather than a link. \\
\cmidrule{3-5}
 &  & Helpful & P6 & Unlike the others, correctly identified the blue linked text in the spam email as a button rather than plain text when given the choice (``Like a button''), though he could not explain why it was blue or what clicking it would do. \\
\midrule
 
\multirow[t]{5}{1.7cm}{DD}
& Language & Unhelpful & P1, P2, P3, P4, P6, P7 & All described DD only by the word ``blocked,'' without a clear sense of what was actually being blocked. P2 inverted its protective function. P4 generalized it into a pop-up blocker. P6 did not know ``Chrome,'' ``downloads,'' or ``history'' at all. P7 read the warning fluently but explained it only by repeating the words on screen, and separately mischaracterized what DD actually does, describing it as ``trying to get your information'' like a phishing attempt rather than a blocked file. \\
\cmidrule{2-5}
 & Information\newline Presented & Unhelpful & P2, P4, P6 & P2 said the notification itself needed to be bigger to be noticed. P4 clicked rapidly and repeatedly on the page before the notification had finished loading, not realizing it took a moment to appear, and only reacted once several had stacked up at once. P6 missed it entirely the second time it appeared and, once shown, agreed it was ``very small.'' \\
 \cmidrule{2-5}
 & Iconography & Helpful & P5 & Independently recognized the octagon-with-X icon and correctly read its meaning (``It's the X,'' i.e., don't do it). \\
 \cmidrule{2-5}
 & Buttons & Unhelpful & P5 & Chose to ``keep'' the flagged download over deleting it, without a clear reason for either option. \\
\cmidrule{2-5}
 & Interactive\newline Elements & Helpful & P1 & Directly interacted with DD's ``Learn more'' link and distinguished it as safe to click, unlike other warning notifications: ``it's okay to, like, click that.'' \\

\midrule
 
\multirow[t]{5}{1.7cm}{NSI}
& Information\newline Presented & Unhelpful & P1, P2, P3, P4, P5, P6 & None engaged with the OS, browser, or full date/time fields without heavy prompting. P1 initially judged both alerts legitimate based on his own dual residency. P2 reversed her judgment repeatedly. P4 misidentified her own browser and OS. P5 did not recognize ``MacOS'' or ``Chrome'' at all. P6 misread ``Madison'' as ``A moon'' and never engaged with the browser or OS fields at all. \\
\cmidrule{3-5}
 & & Helpful & P1, P2, P3, P4, P5, P6, P7 & Location was the detail that worked for all seven once they used it. P1: ``the obvious one is states.'' P2 and P3 both read it immediately. P4 settled on it after initial confusion. P5, once the city was read aloud, connected it to herself immediately. P6 also converged on the city once he looked closely. P7 named both the city and the date/time as the details that mattered to her, correctly identifying the real login by city. Eye-tracking suggested she looked at the date/time far less than she reported using it. \\
\cmidrule{2-5}
 & Interactive\newline Elements & Unhelpful & P2, P4, P5, P6 & P2 and P4 both treated the ``secure your account'' hyperlink as inherently dangerous. P5 correctly identified the blue text as a clickable button, unlike P2 and P4, but declined to find out what it would do. P6 repeatedly tried clicking the non-interactive computer icon and the timestamp, expecting a response; when neither did anything, he reversed his own identity judgment rather than revisit the actual content. \\
\cmidrule{3-5}
 &  & Helpful & P7 & Unlike every other participant, found and used the ``secure my account'' hyperlink correctly and connected it to a real protective action: read the surrounding text, hovered over the link, and clicked through to the password-reset flow. Researchers noted she was the only participant who connected NSI to an action she herself would take. \\
\midrule
 
\multirow[t]{1}{1.7cm}{SRI}
 & Iconography/Color & Unhelpful & P2, P3, P4, P5, P6, P7 & P2 guessed the on-screen time itself was frozen. P3 showed a visual, not functional, preference between two indicator variants. P4 stayed anchored to the weather task. P5 connected it to Wi-Fi. P6 connected it to the clock it was drawn over. P7 showed no recognition when first shown, and even after being told directly what it meant, mixed it up minutes later with the camera indicator, describing SRI as ``somebody's recording you'' through the camera rather than the screen. \\
\midrule
 
\multirow[t]{1}{1.7cm}{MUI}
 & Iconography/Color & Unhelpful & P1, P2, P3, P4, P6, P7 & P1 mapped it to a workplace-availability concept. P2 said its low contrast made it hard to see. P3 and P4 both guessed Wi-Fi/internet status. P6 guessed battery/charger. P7 guessed ``somebody's spying on you,'' a security-flavored guess that may reflect the study's framing as much as an independent read of the icon. \\
\bottomrule
\end{longtable}

\end{document}